\documentclass{emulateapj}

\usepackage{graphics}

\usepackage{enumitem}

\usepackage{amsmath}
\usepackage{xcolor}
\usepackage{soul}

\newcommand{\chandra}{\textit{Chandra }}
\newcommand{\spitzer}{\textit{Spitzer }}

\newcommand{\herschel}{\textit{Herschel }}

\begin{document}


\shortauthors{TEMIM ET AL.}

\shorttitle{Massive Dust Shell in SNR G54.1+0.3} 

\title{A Massive Shell of Supernova-formed dust in SNR G54.1+0.3}

\author{Tea Temim\altaffilmark{1}, Eli Dwek\altaffilmark{2}, Richard G. Arendt\altaffilmark{2,3}, Kazimierz J. Borkowski\altaffilmark{4}, Stephen P. Reynolds\altaffilmark{4}, Patrick Slane\altaffilmark{5}, Joseph D. Gelfand\altaffilmark{6}, and John C. Raymond\altaffilmark{5}}

\altaffiltext{1}{Space Telescope Science Institute, 3700 San Martin Drive, Baltimore, MD 21218, USA}
\altaffiltext{2}{Observational Cosmology Lab, Code 665, NASA Goddard Space Flight Center, Greenbelt, MD 20771, USA}
\altaffiltext{3}{University of Maryland-Baltimore County, Baltimore, MD 21250, USA}
\altaffiltext{4}{North Carolina State University}
\altaffiltext{5}{Harvard-Smithsonian Center for Astrophysics, 60 Garden Street, Cambridge, MA 02138, USA}
\altaffiltext{6}{New York University, Abu Dhabi}

\slugcomment{Accepted by ApJ}

\begin{abstract}

While theoretical dust condensation models predict that most refractory elements produced in core-collapse supernovae (SNe) efficiently condense into dust, a large quantity of dust has so far only been observed in SN~1987A. We present the analysis of \spitzer \textit{Space Telescope}, \herschel \textit{Space Observatory}, Stratospheric Observatory for Infrared Astronomy (SOFIA), and \textit{AKARI} observations of the infrared (IR) shell surrounding the pulsar wind nebula in the supernova remnant G54.1+0.3.
We attribute a distinctive spectral feature at 21~\micron\ to a magnesium silicate grain species that has been invoked in modeling the ejecta-condensed dust in Cas A, which exhibits the same spectral signature. If this species is responsible for producing the observed spectral feature and accounts for a significant fraction of the observed IR continuum, we find that it would be the dominant constituent of the dust in G54.1+0.3, with possible secondary contributions from other compositions, such as carbon, silicate, or alumina grains. The smallest mass of SN-formed dust required by our models is 1.1~$\pm$~0.8~$\rm M_{\odot}$. We discuss how these results may be affected by varying dust grain properties and self-consistent grain heating models.
The spatial distribution of the dust mass and temperature in G54.1+0.3 confirms the scenario in which the SN-formed dust  has not yet been processed by the SN reverse shock and is being heated by stars belonging to a cluster in which the SN progenitor exploded. The dust mass and composition suggest a progenitor mass of 16--27~$\rm M_{\odot}$ and imply a high dust condensation efficiency, similar to that found for Cas A and SN~1987A. The study provides another example of significant dust formation in a Type~IIP SN and sheds light on the properties of pristine SN-condensed dust.

\end{abstract}


\section{INTRODUCTION} \label{intro}

Interstellar dust plays a significant role in virtually all processes governing the evolution of galaxies and is a key ingredient in chemical evolution models and feedback processes important for star formation. Dust can be produced in the stellar wind outflows of massive stars and in the ejecta of core-collapse supernova (SN) explosions. However, the quantity  and relative fraction of dust formed in these sources are still not well understood. 
The question of whether SN explosions are primary sources of dust in the Universe is still under debate. High dust masses observed in high-redshift galaxies suggest that large quantities of dust had to form on timescales of only a few hundred million years \citep[e.g.][]{gall11,valiente11,dwek11,dwek15b,michalowski15}, pointing to core-collapse SNe as the most likely sources.  Dwek, Galliano \& Jones (2009) find that an average SN explosion would need to produce anywhere from 0.1 to 1.0 $\rm M_{\odot}$ of dust in order to explain the observed dust mass of $> 10^8\rm M_{\odot}$ in the galaxy SDSS J1148+5251 at a redshift of 6.4. The range of required masses is strongly dependent on the efficiency of dust destruction in the interstellar medium (ISM) of the galaxy. 

Theoretical dust formation models using classical nucleation theory and the chemical kinematic approach for the formation of molecular precursors do indeed predict that 0.03-0.7 $\rm M_{\odot}$ of dust can form in the ejecta of core-collapse SNe, and that the grain properties and masses depend on factors such as the type of SN explosion, the mass of the SN progenitor, metallicity, and clumping and mixing of the SN ejecta \citep[e.g.][]{todini01,kozasa09,cherchneff09,cherchneff10, sarangi13, sarangi15}. However, observations of supernova remnants (SNRs) and extragalactic SNe in the mid-infrared (IR) have generally revealed dust masses that are orders of magnitude lower, in the range of $10^{-2}-10^{-3}$ $\rm M_{\odot}$ (see Gall et al. 2011 for a review). While it was hypothesized that a large mass of SN dust may be cooler and emitting primarily at far-IR wavelengths, the lack of sensitivity and spatial resolution, in addition to high confusion noise, prohibited clear detections of SN-formed dust in many SNe and SNRs.

\begin{figure*}
\epsscale{0.35} \plotone{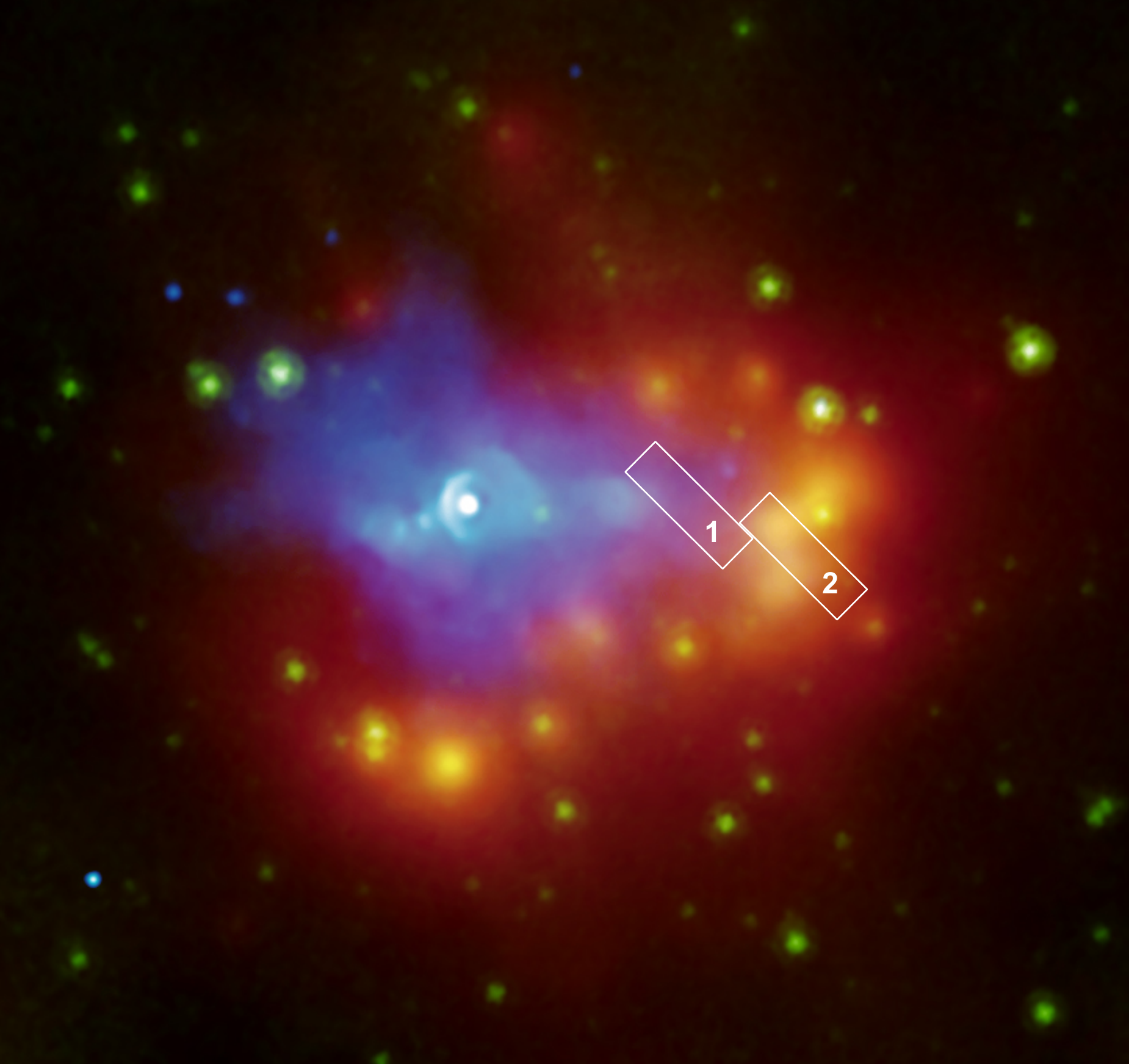}\epsscale{0.35} \plotone{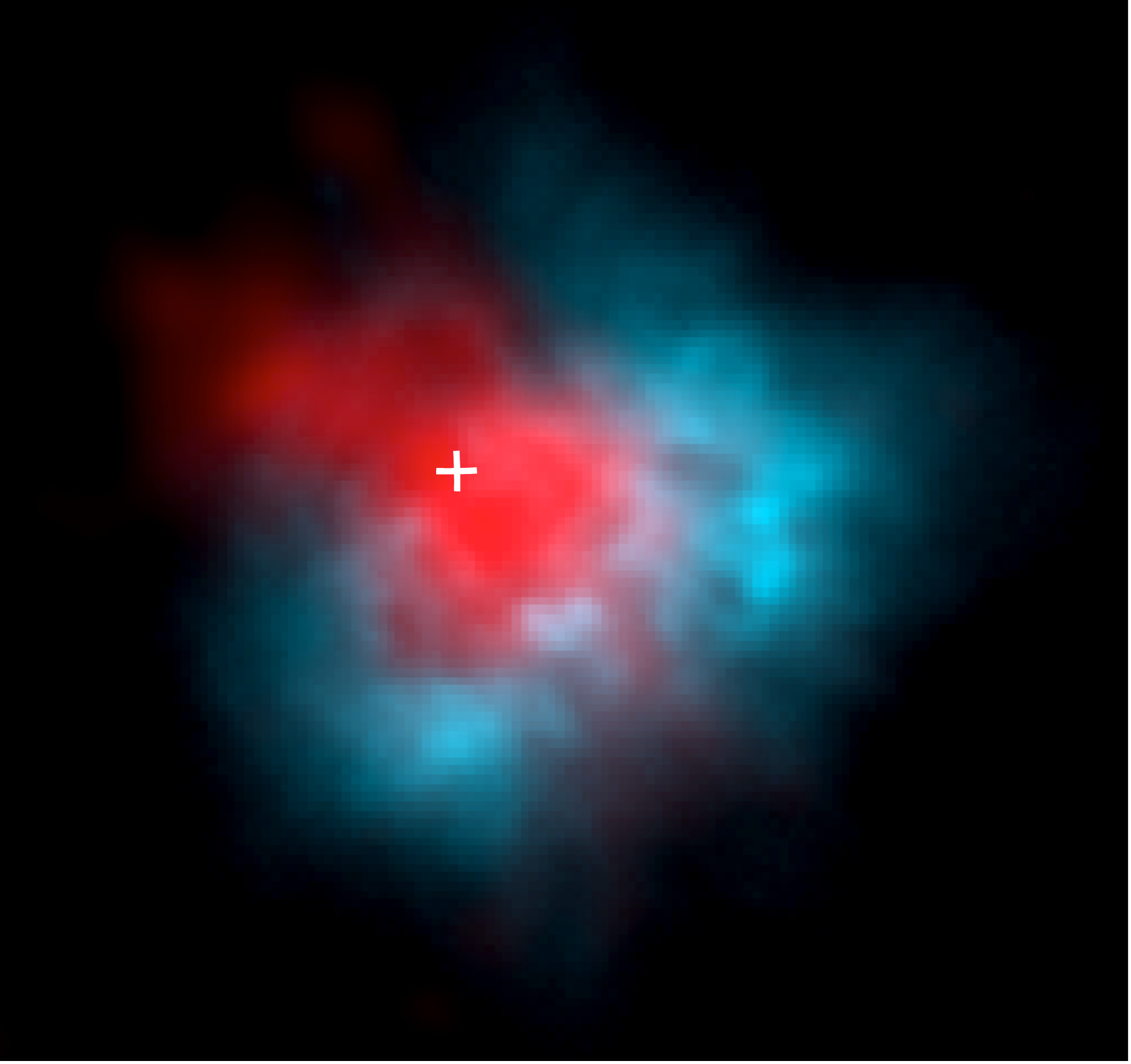}\epsscale{0.35} \plotone{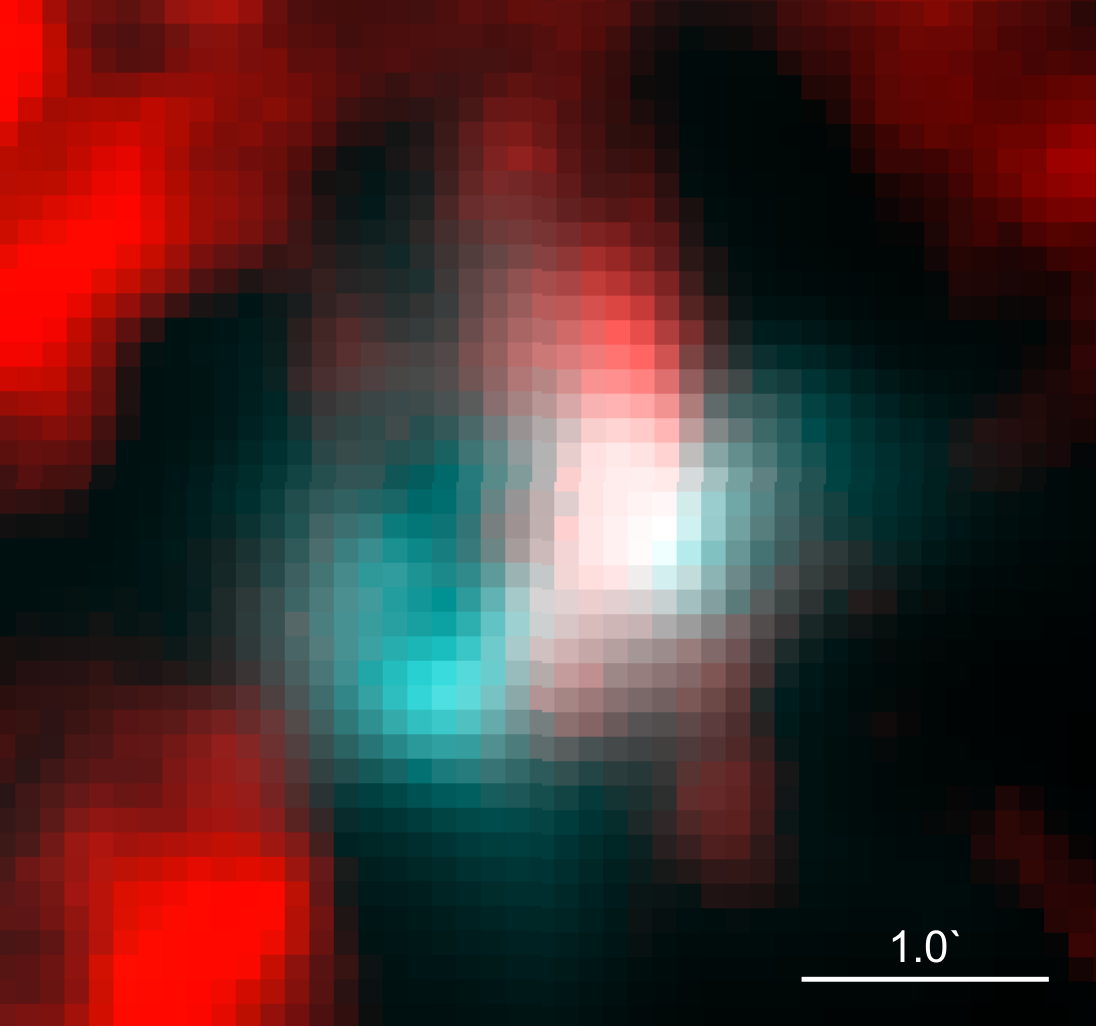}
\caption{\label{3color}\textbf{Left:} A three color composite image of G54.1+0.3 with the \spitzer 8.0 and 24 \micron\ images in green and red, and the \textit{Chandra} X-ray 0.3-10 keV image in blue (NASA/CXC/SAO/\citet{temim10}). The PWN seen in X-rays is surrounded by an IR shell with a radius of 1\farcm3 that emits strongly at 24 \micron. A dozen point sources emitting at both 8 and 24 \micron\ are seen in yellow. The white boxes represent the positions of the high-resolution \spitzer IRS  slits \citep{temim10}, with their corresponding line-subtracted spectra shown in Figure \ref{irsspec}. \textbf{Middle:} A two color image showing the 4.8 Ghz VLA radio synchrotron emission from the PWN in red and \herschel PACS 100 \micron\ infrared emission in cyan. The location of the pulsar is marked by the white plus symbol. \textbf{Right:} \textit{Herschel} SPIRE 250 \micron\ image of the IR shell in red and PACS 70 \micron\ image in teal, convolved to match the 250 \micron\ resolution. The overlay clearly shows the different morphologies of the 70 and 250 \micron\ emission.}
\end{figure*}

Far-IR \herschel observations provided evidence for a significant amount of dust in only three SNRs; the Crab Nebula, Cas A, and SN 1987A. The dust mass in the Crab Nebula is estimated to be between 0.02-0.3 $\rm M_{\odot}$ \citep{gomez12a, temim13, owen15}, while the current mass in Cas A is $\sim0.1$~$\rm M_{\odot}$ \citep{barlow10,sibthorpe10,arendt14}, with 0.8-1.0 $\rm M_{\odot}$ predicted to have formed initially, before being sputtered away by the SN reverse shock \citep{micelotta16}. 

\herschel observations of SN 1987A led to an exciting discovery of a significant mass of dust that likely formed in the SN ejecta \citep{matsuura11}. The ejecta origin was later confirmed with Atacama Large Millimeter/submillimeter Array (ALMA) observations \citep{indebetouw14}. The latest estimate indicates that at least $\sim$ 0.5 $\rm M_{\odot}$ of dust formed in the explosion. However, the composition of the dust is still a matter of debate \citep{matsuura15,dwek15,wesson15}, and how much will eventually survive the passage of the SN reverse shock is still unknown.

In this paper, we present a follow-up study of the dusty shell in the SNR G54.1+0.3, indicating that this system contains a significant mass of SN-formed dust. G54.1+0.3 contains a pulsar wind nebula (PWN) with a well-defined jet and torus structure, exhibiting properties similar to the Crab Nebula \citep{lu01,temim10}. The age of the system is estimated to be 1500--3000 yrs \citep{chevalier05,bocchino10,gelfand15}. Based on the spectral types of stars in the surrounding stellar cluster, \citet{kim13} constrain the range of the progenitor mass to be 18--35 $\rm M_{\odot}$ that likely resulted in a Type IIP SN explosion. This range overlaps with the results of \citet{gelfand15} who derived a progenitor mass of 15--20 $\rm M_{\odot}$ using a model for the dynamical and radiative evolution of a PWN inside an SNR.

The \spitzer mid-IR images of this system revealed a shell of emission surrounding the PWN and a dozen point sources embedded in the shell with an apparent 24 \micron\ IR excess. Due to this excess, the point sources were originally attributed to young stellar objects (YSOs) whose formation was triggered by the progenitor star \citep{koo08}. However, the analysis of line emission detected with the Infrared Spectrograph (IRS) aboard \spitzer\ revealed that the shell has a high expansion velocity of up to 500 $\rm km \: s^{-1}$ and that the emission likely originates from SN ejecta. The IR excess in the point sources can be attributed to SN-formed dust that is heated to higher temperatures as it blows past early-type stars that are members of the cluster in which the SN exploded \citep{temim10}. \citet{kim13} used near-IR spectroscopic observations of the stellar sources to determine their spectral types and concluded that they are indeed late O- or early B-type stars that show no evidence for emission lines often present in Herbig Ae/Be stars. 
 \citet{kim13} also also constrained the distance to G54.1+0.3 by calculating the photometric distances to the stars in the shell based on their  spectral energy distribution (SED)-fitted temperatures. Their derived distance range is 4.6--8.1~kpc, with an average of 6.0~$\pm$~0.4~kpc.
 
In this paper, we analyze the mid- and far-IR emission from the shell surrounding the PWN in G54.1+0.3 using the \spitzer \textit{Space Telescope}, \herschel \textit{Space Observatory}, Stratospheric Observatory for Infrared Astronomy (SOFIA), and \textit{AKARI} observations in order to confirm the origin and determine the properties of the dust in the IR shell. 

\begin{deluxetable}{lccc}
\tablecolumns{4} \tablewidth{0pc} \tablecaption{\label{tab1}OBSERVED INFRARED FLUX DENSITIES}
\tablehead{
\colhead{Instrument} & \colhead{Wavelength} & \colhead{Flux Density} & \colhead{Extinction} \\
\colhead{}  & \colhead{($\micron$)} & \colhead{(Jy)} & \colhead{Correction}
}

\startdata
\textit{AKARI} IRC & 15.6 & 2.6 $\pm$ 0.26 & 1.38 \\
SOFIA FORCAST & 19.7 & 11.9 $\pm$ 2.4  & 1.38  \\
\spitzer MIPS & 24.0 & 23.1 $\pm$ 2.1  & 1.24  \\
SOFIA FORCAST& 25.3 & 12.7 $\pm$ 2.5 & 1.13 \\
SOFIA FORCAST& 31.5 & 42.5 $\pm$ 8.5 & 1.12 \\
SOFIA FORCAST & 34.8 & 41.7 $\pm$ 8.3 & 1.11 \\
\herschel PACS & 70.0 & 87.9 $\pm$ 11.4 & \nodata  \\
\herschel PACS  & 100 &  68.8  $\pm$ 13.4 & \nodata \\
\herschel PACS & 160 &  29.0 $\pm$  14.9 & \nodata \\ 
\herschel SPIRE & 250 & 6.9  $\pm$ 5.2 & \nodata  \\
\herschel SPIRE & 350 & 1.6 $\pm$ 2.8 & \nodata \\
\herschel SPIRE & 500 &  0.4 $\pm$ 1.1 & \nodata

\enddata
\tablecomments{Measured background-subtracted infrared flux densities for the G54.1+0.3 shell before extinction correction. The listed extinction correction factors based on \citet{xue16} were used in the SED fitting. See Section~\ref{sofia} for more on the discrepancy between the measured MIPS 24 \micron\ and FORCAST 25 \micron\ flux densities.
}
\end{deluxetable}



\section{OBSERVATIONS AND DATA REDUCTION} \label{obsv}

\subsection{\textit{Spitzer} Observations}

This work includes \textit{Spitzer} imaging and spectroscopy that were previously analyzed and presented in \citet{temim10}. 
We include the Multiband Imaging Photometer (MIPS) 24 \micron\ image taken on 2005 May 15 under the program ID 3647 (PI: Slane), and the \spitzer Infrared Spectrograph \citep[IRS;][]{houck04} high-resolution spectra taken at two different positions on the IR shell that are indicated by the white boxes in Figure~\ref{3color}. Position 1 is located at the interface of the IR shell and the pulsar's jet seen in X-rays, while Position 2 is located at the brightest peak in the 24 \micron\ image that \citet{temim10} call the ``IR knot". The reduction of the \textit{Spitzer} observations is discussed in detail in \citet{temim10}.

\subsection{\textit{Herschel} Observations}

\herschel imaging of G54.1+0.3 was obtained with the Photodetector Array Camera \citep[PACS;][]{poglitsch10} at 70, 100, 160 \micron\, and the Spectral and Photometric Imaging Receiver \citep[SPIRE;][]{griffin10} at 250, 350, and 500 \micron. The PACS observations were performed on 2011 Nov 5 using the ``scan map'' mode and a scan speed of 20 arcsec $\rm s^{-1}$ (proposal ID: \textit{OT1\_ttemim\_1}, observations IDs: 1342231919-1342231922). The SPIRE observations were performed on 2011 May 2 and 2011 Oct 23 in the ``parallel mode", under the proposal ID \textit{KPOT\_smolinar\_1} (observation IDs: 1342219812, 1342219813, 1342231341, and 1342231342). The imaging observations were processed and reduced with the \textit{Herschel} Interactive Processing Environment \citep[HIPE;][]{ott10} version 14.0.0 and images produced using the \textit{MADmap} software \citep{cantalupo09}. The resulting PACS and SPIRE images are in units of Jy/pixel with pixel scales (full width at half-maxima, FWHM) of 1.6, 1.6, 3.2, 6, 10, and 14\arcsec/pixel (6, 8, 12, 18.1, 24.9, and 36.4\arcsec) for the 70, 100, 160, 250, 350, and 500 \micron\ images, respectively. The calibration uncertainties for the PACS and SPIRE images are assumed to be 10\% and 7\%, respectively.

\herschel spectra were obtained with the PACS Integral Field Unit (IFU) Spectrometer \citep{poglitsch10} in the ``range spectroscopy" mode, covering the lines of 63.18 and 145.53 \micron\ [\ion{O}{1}], 88.36 \micron\ [\ion{O}{3}], and 157.74 \micron\ [\ion{C}{2}]. The IFU has 5 $\times$ 5 spaxels, measuring 9\farcs4 on a side, and the G54.1+0.3 shell was mapped in nine pointings (3 $\times$ 3 grid), with an additional pointing for the off-source background. The level 2 data were analyzed using HIPE version 14.0.0. The observations were used to determine whether the line emission significantly contributes to the integrated fluxes measured from the \herschel images.

\begin{figure*}
\center
\epsscale{1.08} \plotone{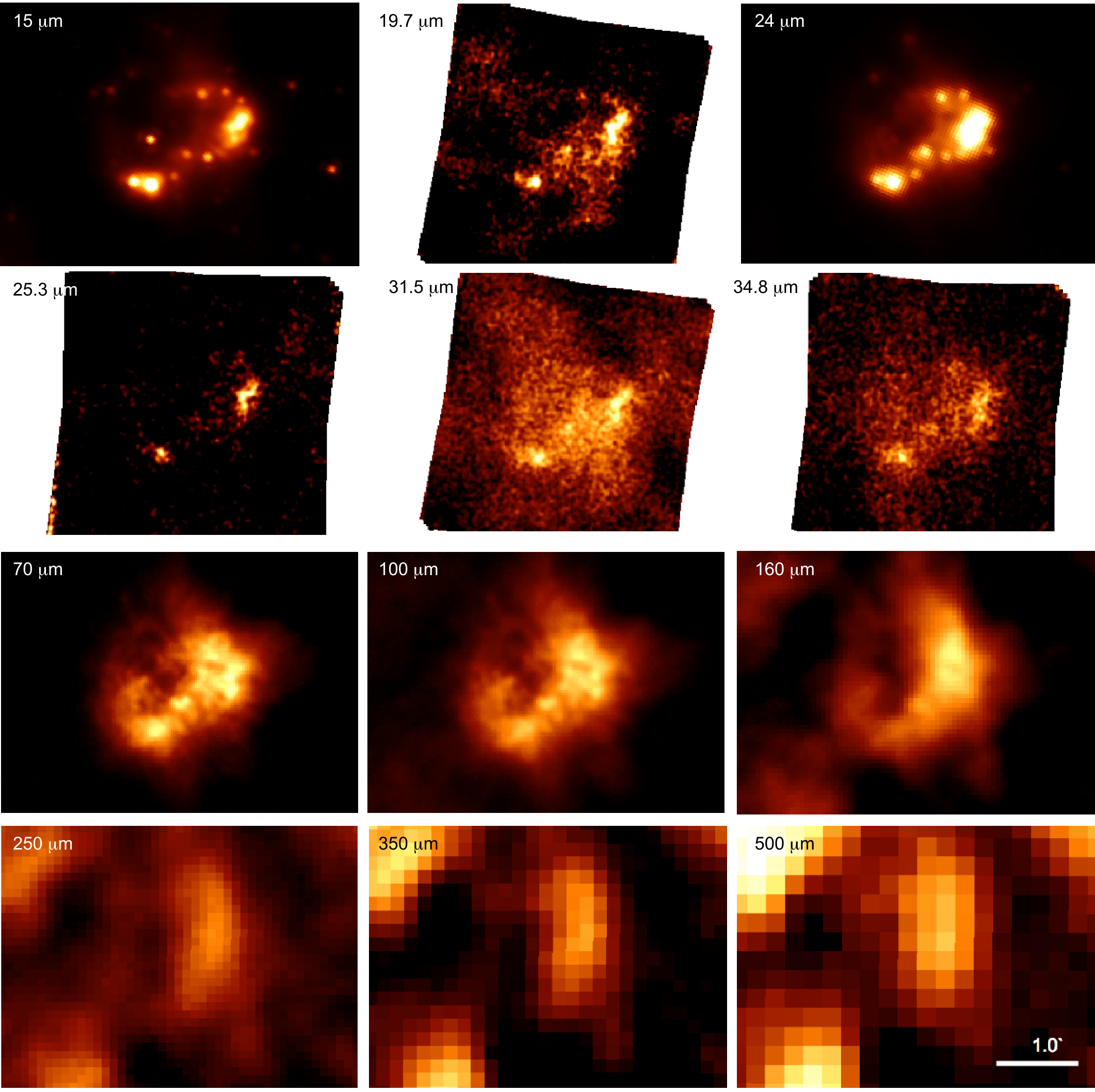}
\caption{\label{imgpanel}Imaging observations of the infrared shell in G54.1+0.3, including the AKARI 15 \micron, SOFIA 19.7, 25.3, 31.5, and 34.8 \micron, \spitzer MIPS 24 \micron \citep{temim10}, \herschel PACS 70, 100, and 160 \micron, and \herschel SPIRE 250, 350, and 500 \micron\ images. All images are shown on a linear color scale. The integrated background-subtracted flux densities of G54.1+0.3 are listed in Table~\ref{tab1}.}
\end{figure*}

\subsection{SOFIA Imaging} \label{sofia}

G54.1+0.3 was observed by SOFIA on 2016 Feb 18 (OC4-A Flight 8), using the Faint Object infraRed
CAmera for the SOFIA Telescope (FORCAST). 
Simultaneous imaging in the
short and long wavelength channels, using a cold dichroic beamsplitter,
was done in F197 (+F315) and F253 (+F348) broadband filters with
effective wavelengths of 19.7 (31.5), and 25.3 (34.8) $\mu$m and bandwidths of
5.5 (1.86), 5.7 (3.8) $\mu$m, respectively.
G54.1+0.3 is an extended
($2\farcm$5 in size) IR source that fits nicely into the
$3\farcm 4 \times 3\farcm 2$ FORCAST field-of-view. Chopping and nodding
were done symmetrically about the telescope's optical axis, using symmetric
nod-match-chop (NMC) with a $2\farcm5$ chop throw (the nod throw is matched to
the chop throw in this chopping and nodding mode). This fairly large chop throw
degrades the image quality by asymmetric smearing of the PSF at a level of
$2\arcsec$ per $1\arcmin$ of chop amplitude. The resulting image degradation
at the long wavelengths (31.5 $\mu$m and 34.8 $\mu$m) is quite modest, but at
short wavelengths it becomes noticeable for unresolved point sources.
No image dithering was employed. After postprocessing, the image pixel is
$0\farcs 768$. Data reduction was done at the SOFIA Science Center \citep[see][for a thorough description of data acquisition and reduction for FORCAST observations]{herter13}. We use the processed, flux calibrated
(Level~3) images obtained from the raw (Level~1) observations of G54.1+0.3
using the standard FORCAST processing pipeline. The total exposure times are
1625, 1750, 1875, and 1670 s in F197, F253, F315, and F348 filters,
respectively. The calibration uncertainty for each band is assumed to be $\sim$~20\% \citep{herter13}.
The measured flux of the 25.3 \micron\ image in particular is significantly lower than the measured MIPS 24 \micron\ flux. We suspect that this observation was not sensitive to the fainter extended emission due to the poor throughput of the FORCAST 25.3 \micron\ filter. The lack of faint extended emission is also evident in the FORCAST 25.3 \micron\ image shown in Figure~\ref{imgpanel}. Due to the unreliable flux estimate for the extended emission, we exclude the 25.3 \micron\ SOFIA data point from our analysis and fitting of the IR SED. 

\subsection{\textit{AKARI} Image}

In our analysis, we also included the \textit{AKARI} Infrared Camera (IRC) 15 \micron\ image of G54.1+0.3 taken on 2007 April 17 (ID: 1401070 001) that has an angular resolution of 5\farcs7 and was processed with standard pipeline processing. The image was previously presented by \citet{koo08}.

\section{General Morphology}\label{morph}

The individual IR images of G54.1+0.3 from 15--500 \micron\ are shown in Figure~\ref{imgpanel}, while the morphological comparison of X-ray, radio, and IR wavelengths is shown in the color images of Figure~\ref{3color}. The left panel of Figure~\ref{3color} shows the three-color image, with the \chandra X-ray emission in blue, and \spitzer 8 and 24 \micron\ emission in green and red. The X-ray image clearly shows the location of the pulsar (white point source), surrounded by a PWN with a well defined torus structure \citep{lu02,temim10}. There is a jet extending from the pulsar directly towards the west. The 8 \micron\ emission in green primarily shows emission from stellar sources. The 24 \micron\ image shows that the PWN is surrounded by a shell of IR emission, approximately 1\farcm3 in radius. The dozen point sources that appear yellow in the image also have strong 24 \micron\ fluxes. This is more clearly seen in the individual 24 \micron\ image in Figure~\ref{imgpanel}. The sources appear to be embedded in the diffuse IR emission and arranged in a ring-like structure.

The middle panel of Figure~\ref{3color} shows the PWN's 4.8~GHz radio emission from the Very Large Array (VLA) in red \citep{velusamy88}, and \herschel PACS 100 \micron\ emission in cyan. The location of the pulsar is marked by the white plus symbol. The PWN has a similar spatial extent at radio and X-ray wavelengths. The radio nebula fills the cavity of the shell that is traced by the 100 \micron\ emission and appears to have expanded into this material. The complementary morphology of the radio and IR emission confirms the association and possible interaction between the PWN and the IR shell.

The 15 and 24 \micron\ images show point-like emission that coincides with the stellar sources observed at 8 \micron\ (see Figure~\ref{imgpanel}). In the higher resolution SOFIA images, these sources appear to be somewhat more extended. While the 31.5--100 \micron\ images do show some structure and regions of enhanced emission, there is no clear evidence for point-like emission. At 160 \micron\ and above, the shell emission appears more uniform, partly due to the lower resolution at these wavelengths. While a full shell is still apparent in the 160 and 250 \micron\ images, the emission at 350 and 500 \micron\ primarily arises from a more localized region in the western part of the shell. This may suggest that there is either a higher fraction of colder dust present in this region of the shell, leading to a higher flux density at 500 \micron, or that there is more mass, and therefore higher emission, concentrated in the western region. The shell morphology in the mid- and far-IR appears to be different. The emission at 70 and even at 100 \micron\ appears more circular, while the shell structure at 250 \micron\ appears more elongated in the north/south direction. This is most apparent in the two-color image shown in the third panel Figure~\ref{3color}, in which the 70 \micron\ image shown in teal has been convolved to the resolution of the 250 \micron\ image shown in red. The 160 \micron\ emission in Figure~\ref{imgpanel} seems to be a blend of these two morphologies. In later sections, we will show that the emission in the far-IR either arises from distinct dust component than the one producing the mid-IR emission, or that it is associated with background emission.

\begin{figure*}
\center
\epsscale{0.45} \plotone{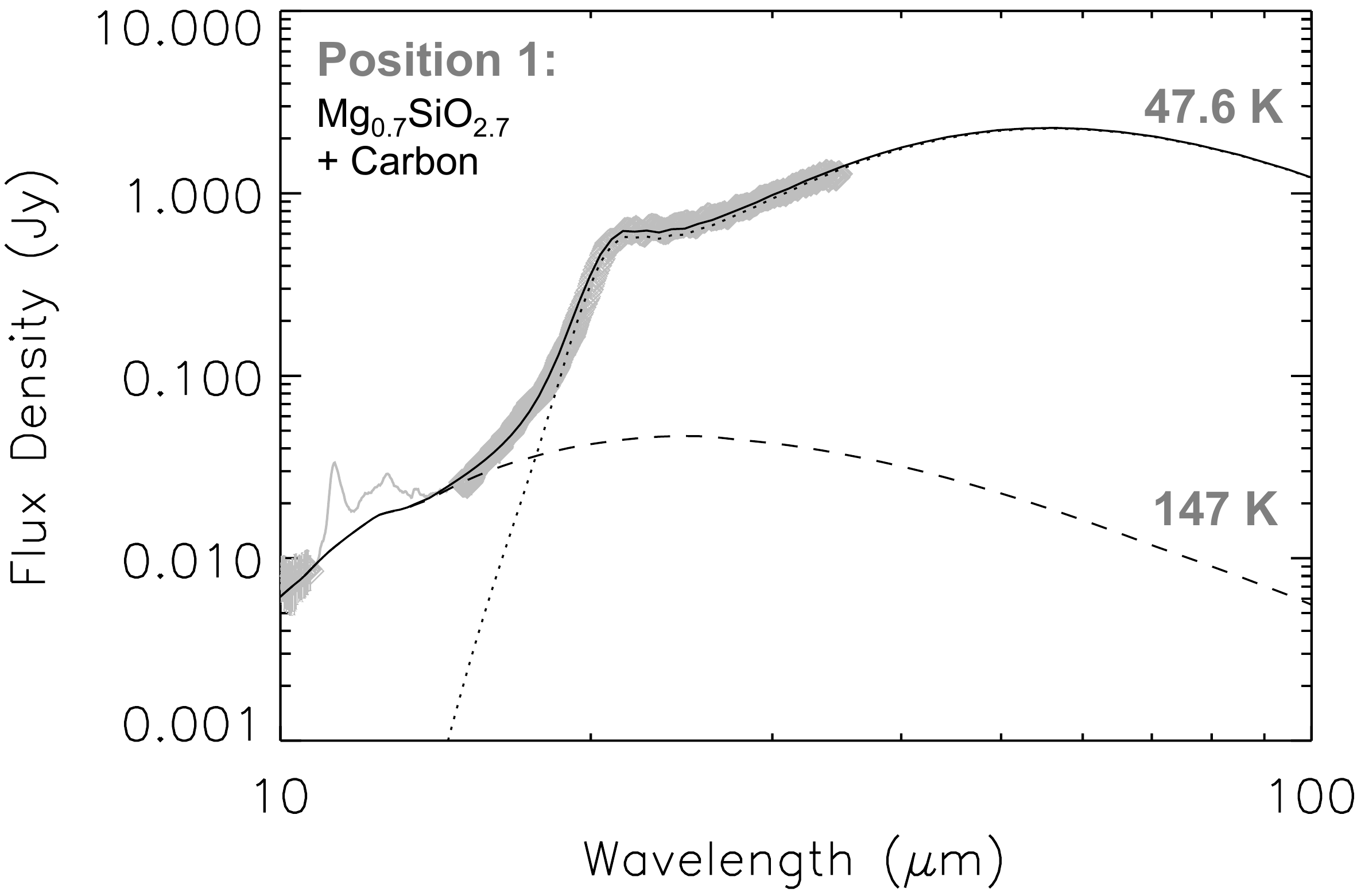}
\epsscale{0.45} \plotone{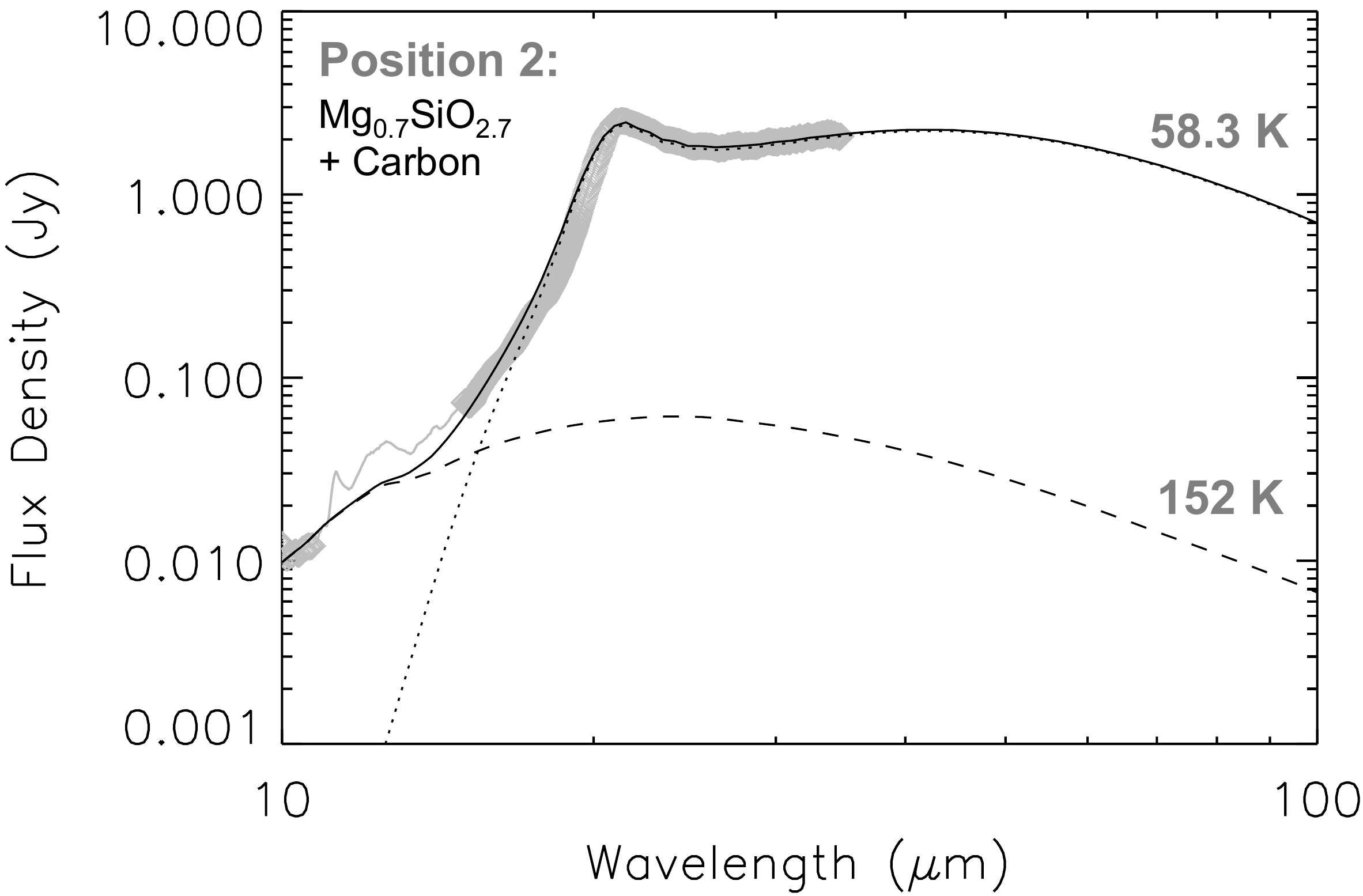}
\epsscale{0.45} \plotone{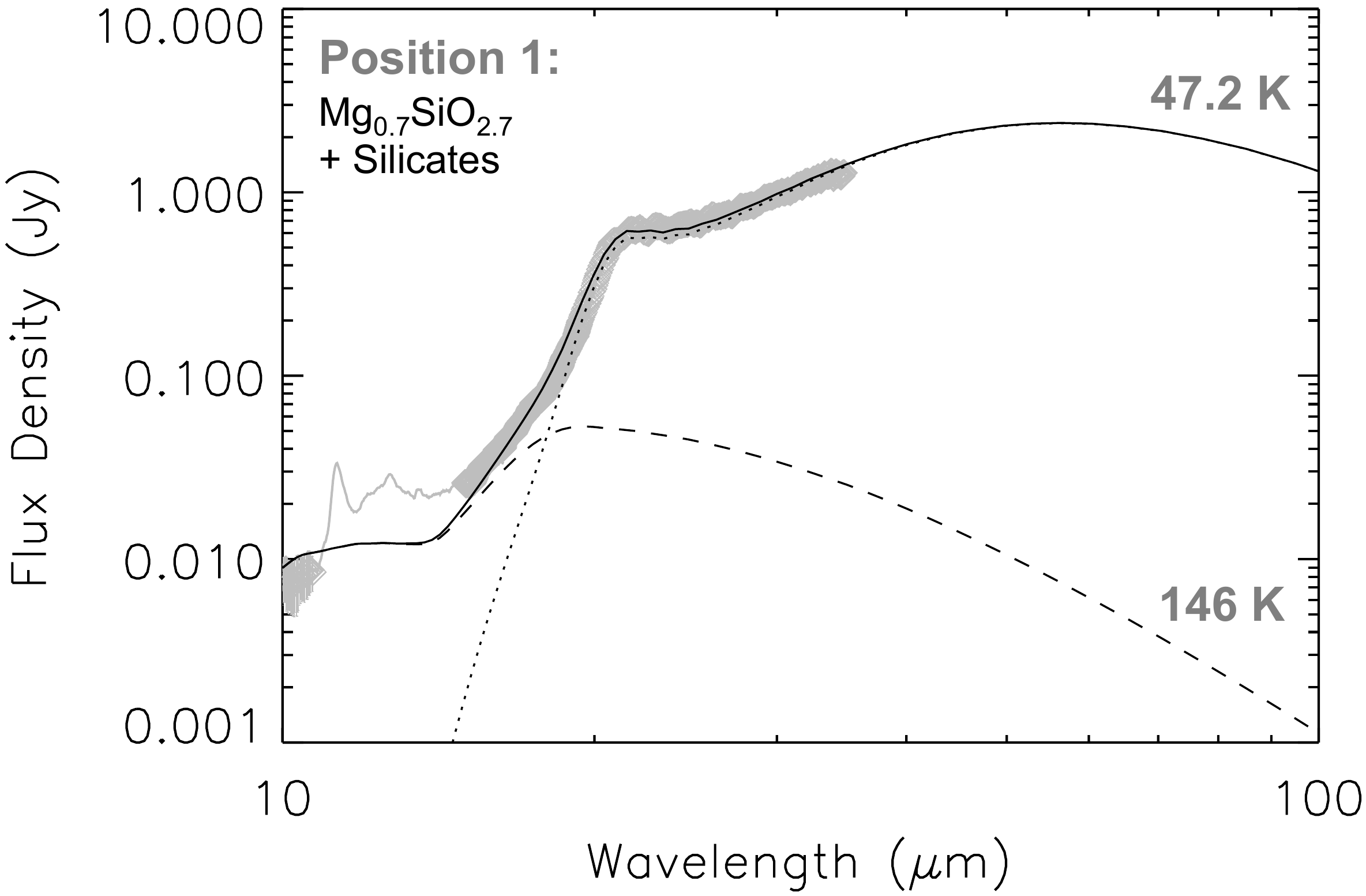}
\epsscale{0.45} \plotone{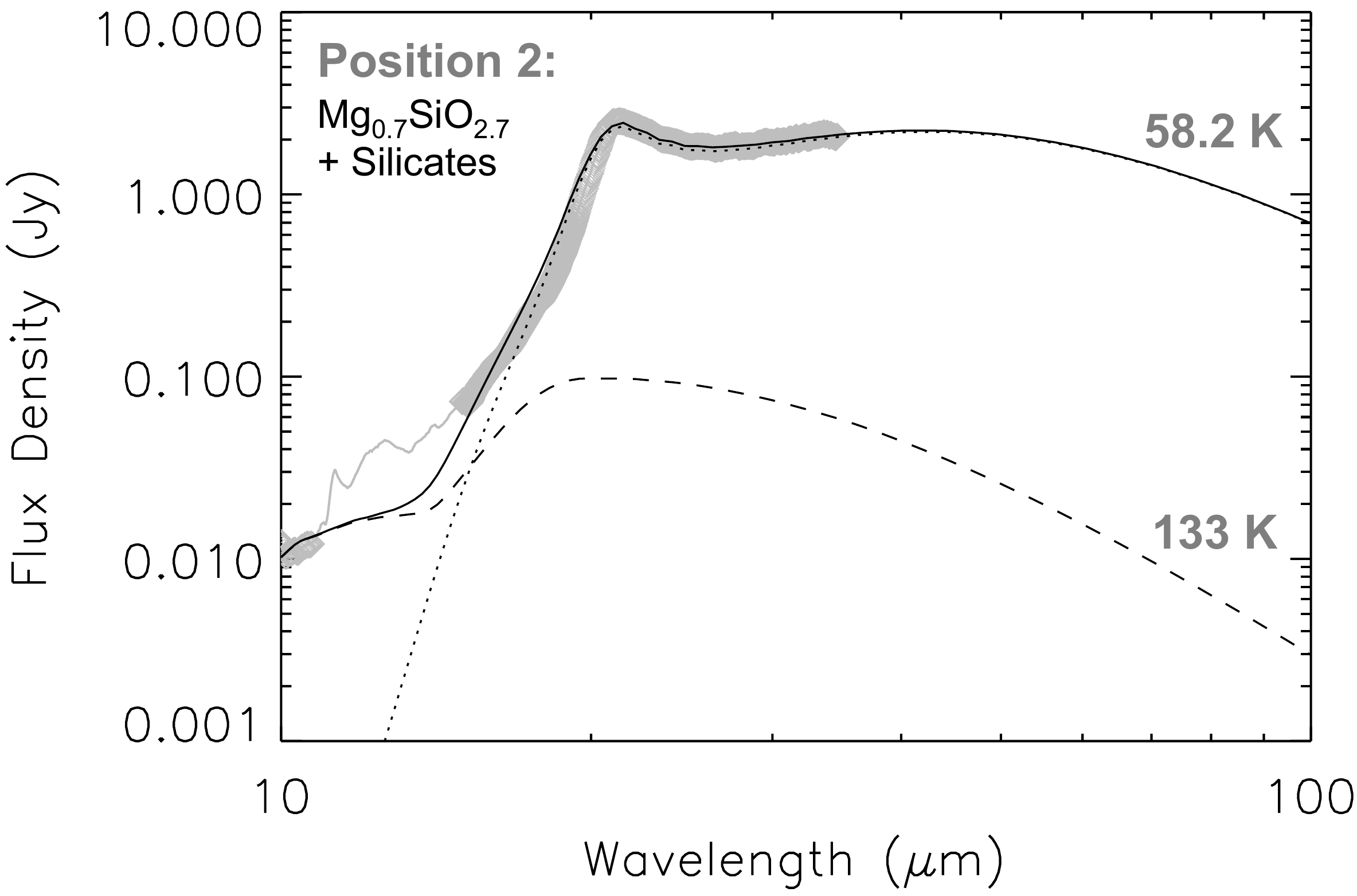}
\caption{\label{irsspec}The high-resolution, line-subtracted \spitzer IRS spectra from two positions shown in Figure \ref{3color} fitted with the $\rm Mg_{0.7}SiO_{2.7}$ dust component shown by the dotted curves, plus a carbon (silicate) component in the top (bottom) row shown as the dashed curves. The gray bands represent the statistical uncertainties in the spectrum. The wavelength region between 11 and 13 \micron\ represented by the gray solid line contains PAH features and was excluded from the fit. The fits are consistent with a hotter carbon or silicate dust component emitting at a temperature of $\sim$~150 K and $\sim$~140 K, respectively, plus a $\rm Mg_{0.7}SiO_{2.7}$ component emitting at $\sim$~47~K at position 1 and  $\sim$~58~K at position 2. The different shapes of the IR spectra at the two positions can be explained by variations in the temperature of the $\rm Mg_{0.7}SiO_{2.7}$ dust component.}
\end{figure*}

\begin{figure}
\center
\epsscale{1.15} \plotone{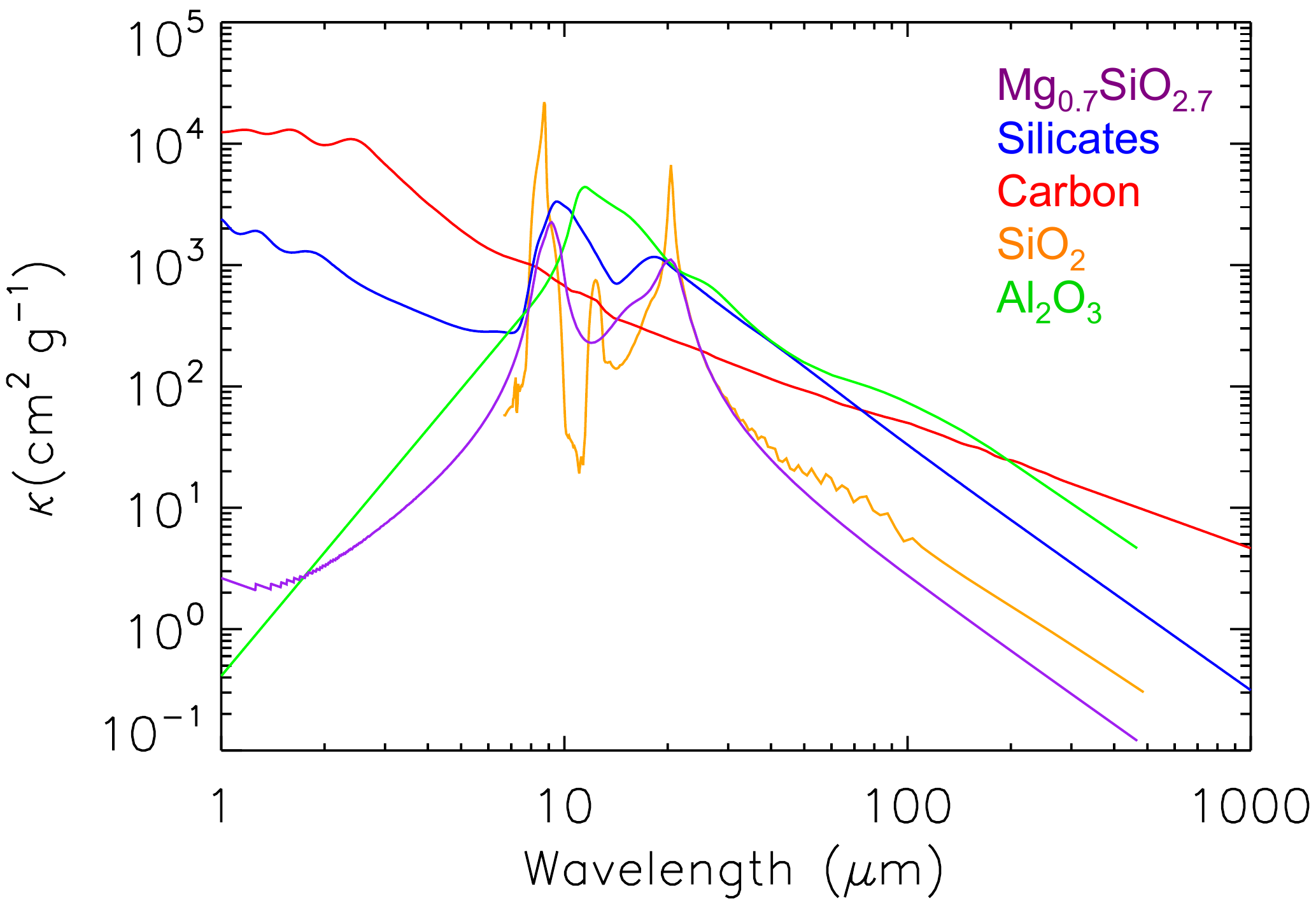}
\caption{\label{kappas}Mass absorption coefficients for the dust species used in the fits, with $\rm Mg_{0.7}SiO_{2.7}$ in purple \citep{jager03}, silicates in blue \citep{weingartner01}, amorphous carbon in red \citep{rouleau91}, $\rm SiO_2$ in orange \citep{henning97}, and $\rm Al_2O_3$ in green \citep{begemann97}.}
\end{figure}

\section{Origin of the IR Shell}\label{origin}

The study of the \spitzer IR imaging and spectroscopy of the shell in G54.1+0.3 concluded that the emission most likely arises from newly formed dust in the SN ejecta \citep{temim10}. In this scenario, the IR emitting material is the inner SN ejecta that the PWN is expanding into and that has not yet been reached by the SNR reverse shock. 
In addition to the complementary morphologies of the PWN and the shell that suggest that they are interacting, modeling of the observed IR line intensities from the shell implies that the PWN is driving a 25-30 $\rm km\:s^{-1}$ shock into the ejecta. In one particular location (coinciding with Region~2 in Figure~\ref{3color}), the measured gas density is as high as 500--1300 $\rm cm^{-3}$. This region coincides with a brightness peak in the MIPS 24~\micron\ image and may represent ejecta material that has been compressed by the shock driven by the pulsar's jet \citep{temim10}.
Another piece of evidence that suggests that the shell material is composed of SN ejecta is the composition of the dust itself (see Section~\ref{comp}). 

\citet{temim10} argued that the IR excess in the point sources in the \spitzer 24 \micron\ image (see Figures~\ref{3color} and \ref{imgpanel}) is not stellar in origin, but that the emission instead arises from ejecta dust that is being radiatively heated by early-type stars belonging to a stellar cluster in which the SN exploded. The stars heat the dust in their immediate vicinity to higher temperatures, giving rise to a mid-IR excess that appears point-like due to the limited spatial resolution. 
This scenario is consistent with the morphology of the images in Figure~\ref{imgpanel}, in which we see that the regions with enhanced brightness in the long wavelength SOFIA and the MIPS 70 \micron\ images and longward do not appear point-like, but are instead more extended. \citet{kim13} classified the spectral type of each star in the shell and concluded that they are indeed late O- and early B-type stars with no evidence for any emission lines that are typically associated with Herbig Ae/Be stars. This provides further evidence that the 24 \micron\ point-like emission is not intrinsic to the stars.

The evidence outlined above strongly suggests that the IR shell is not circumstellar in origin and that the mid-IR point sources are not YSOs. An additional argument against a circumstellar origin is the lack of any X-ray thermal emission that would results from the interaction of the SN blast wave with the surrounding dense material. Additionally, a reverse shock would have formed as the SN blast wave encountered the cloud and significantly disrupted the PWN. However, the well-defined torus and jet structures and the modeling of the PWN evolution \citep{gelfand15} strongly suggest that this has not occurred and that the material that we observe in the shell arises from inner SN ejecta.

The G54.1+0.3 system is analogous to the Crab Nebula, which also consists of a PWN expanding into and radiatively heating SN ejecta and dust \citep[e.g.][]{hester08}. However, in the case of G54.1+0.3, the stars that are part of the cluster in which the progenitor was born serve as the primary heating sources for the SN-condensed dust as it blows past them. 
We note that additional ejecta material may be present beyond the radius of the observed IR shell, but remain undetected due to its cold temperature.

\begin{figure*}
\center
\epsscale{0.37} \plotone{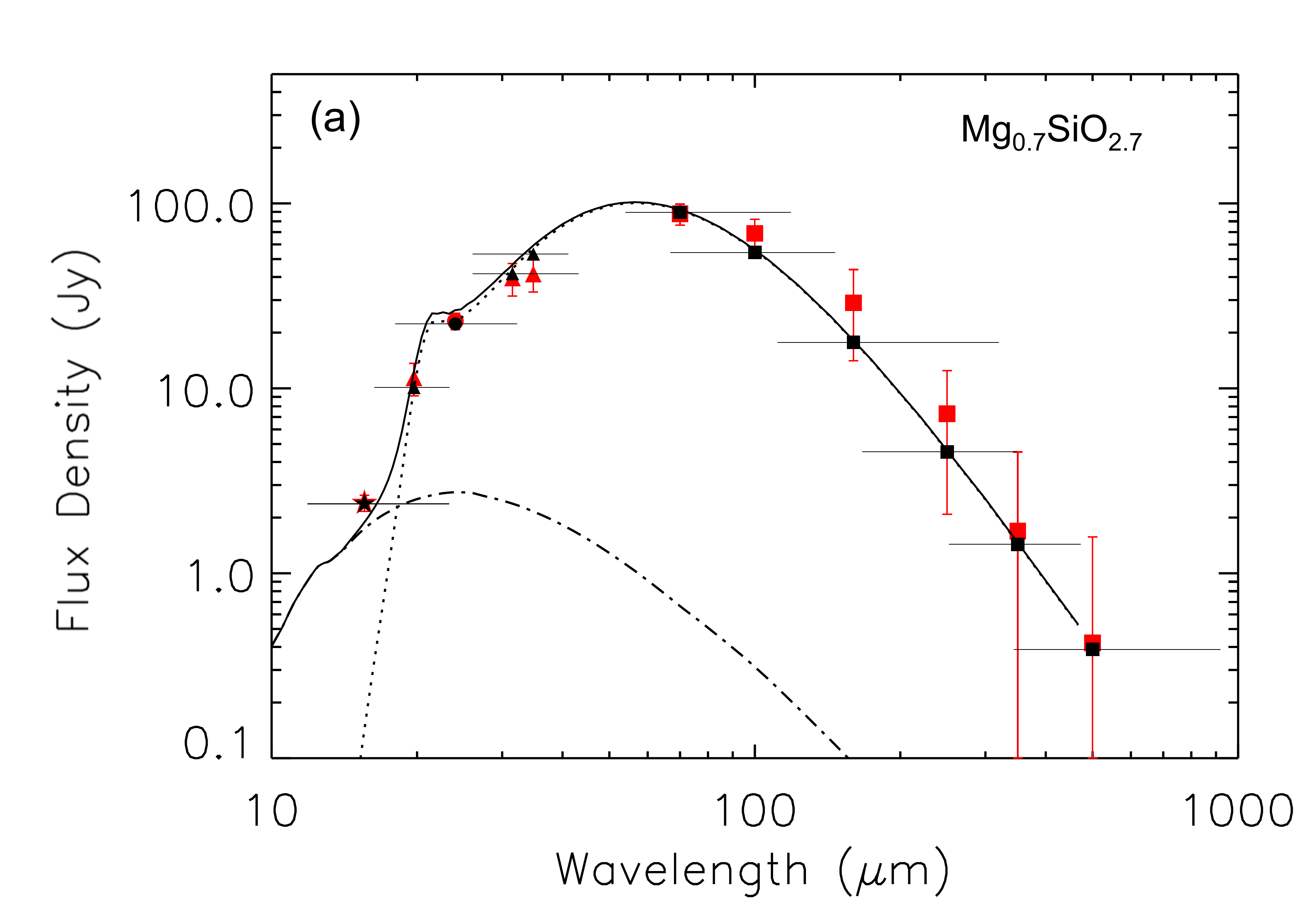}
\epsscale{0.37} \plotone{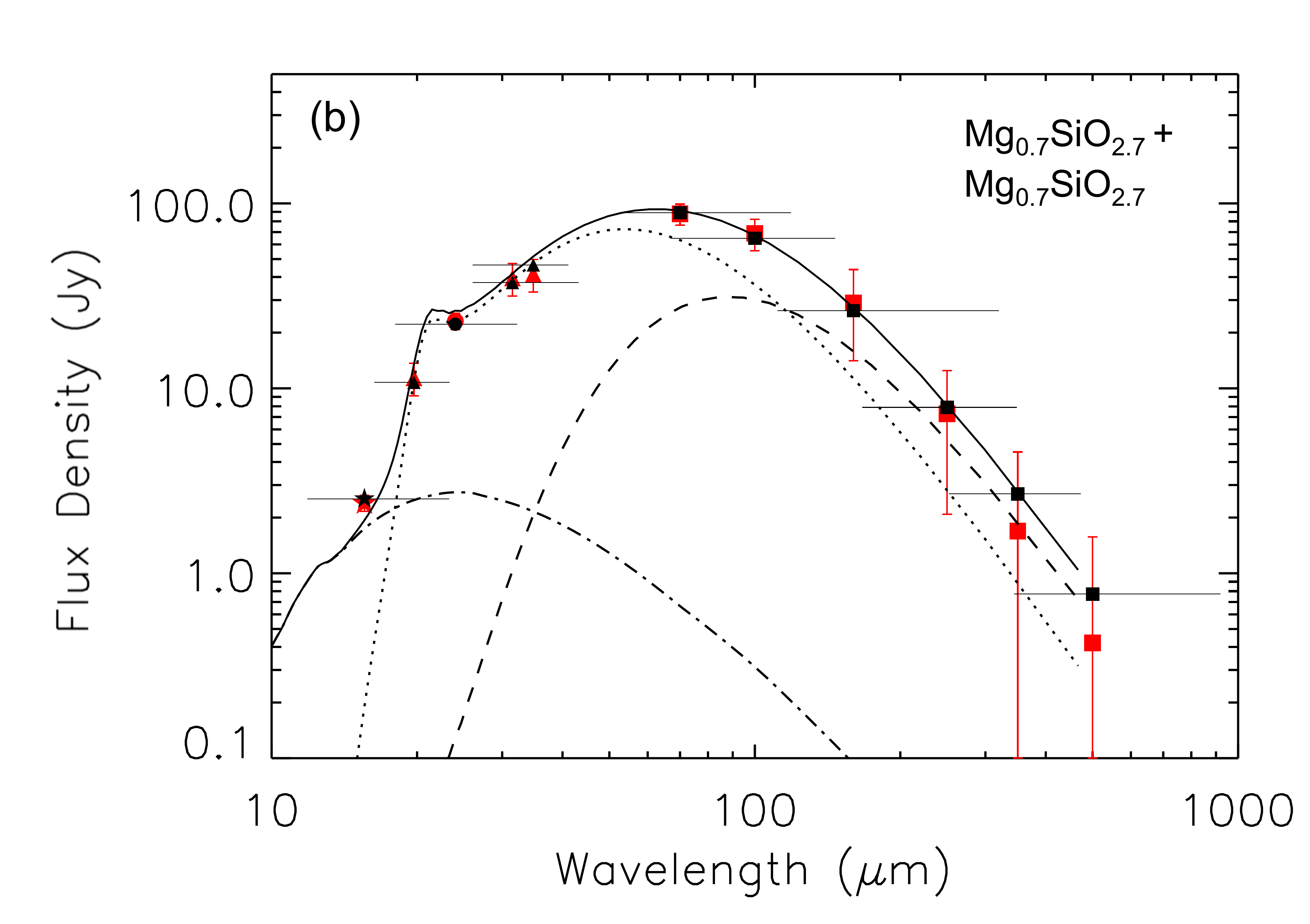}
\epsscale{0.37} \plotone{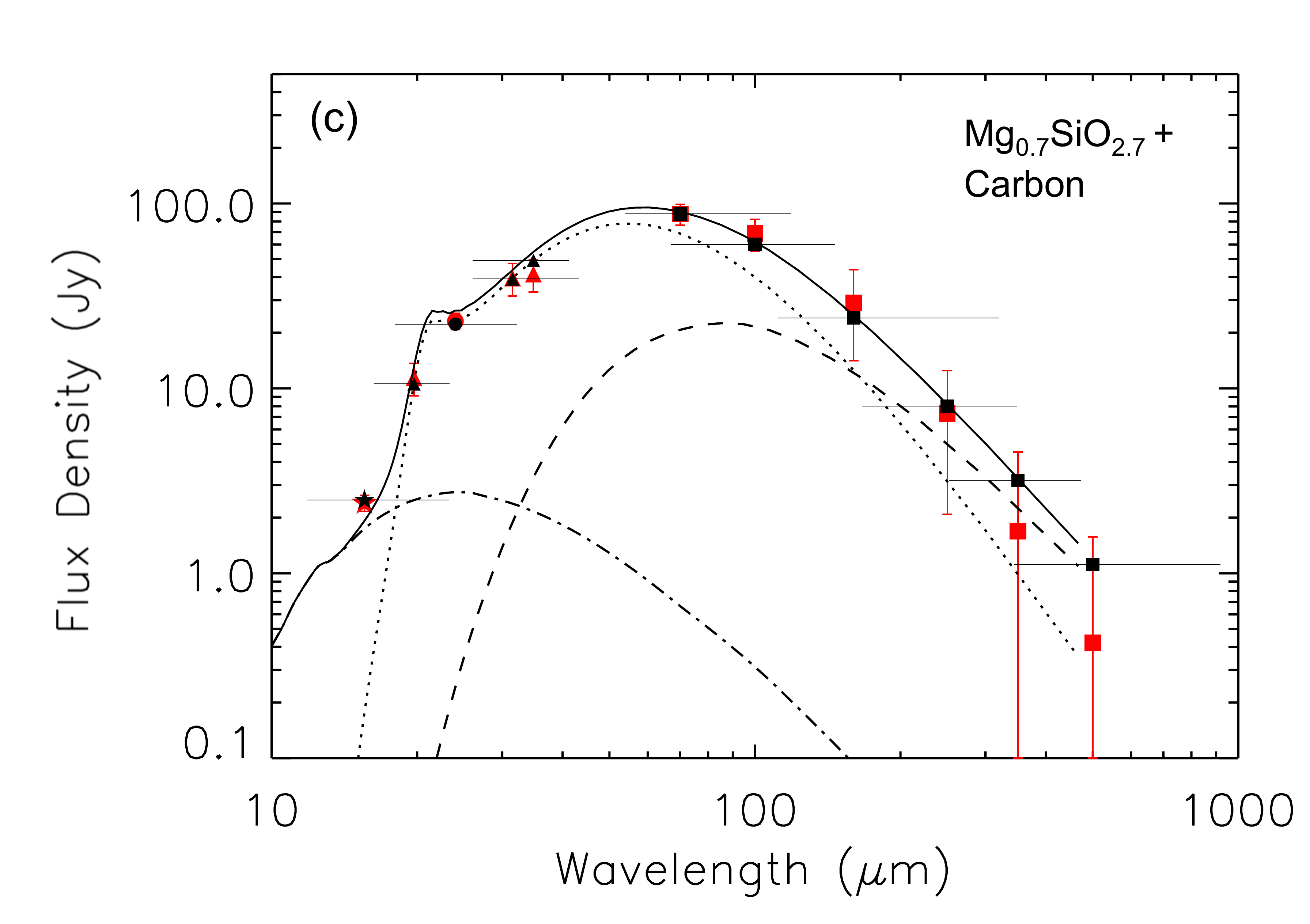}
\epsscale{0.37} \plotone{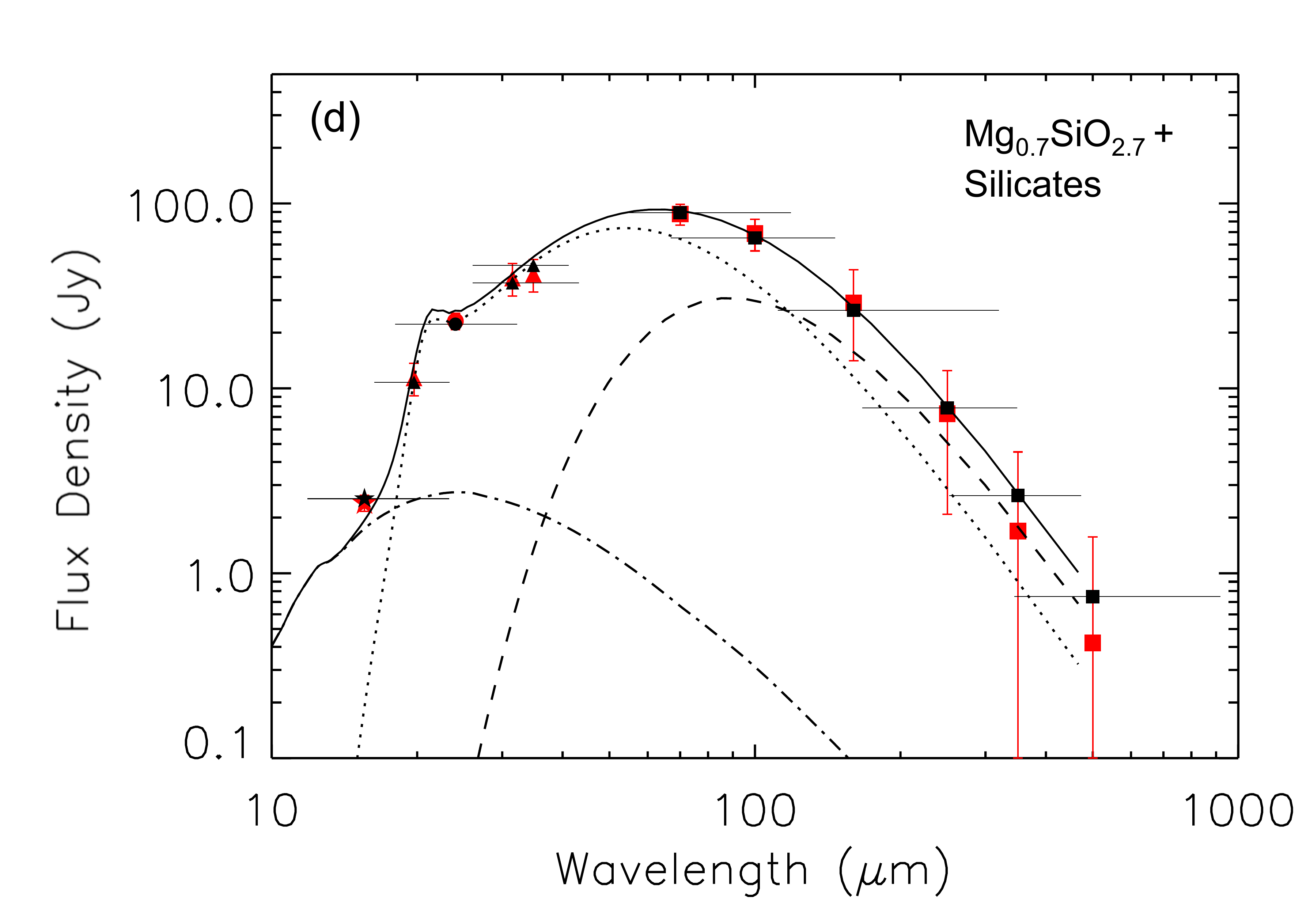}
\epsscale{0.37} \plotone{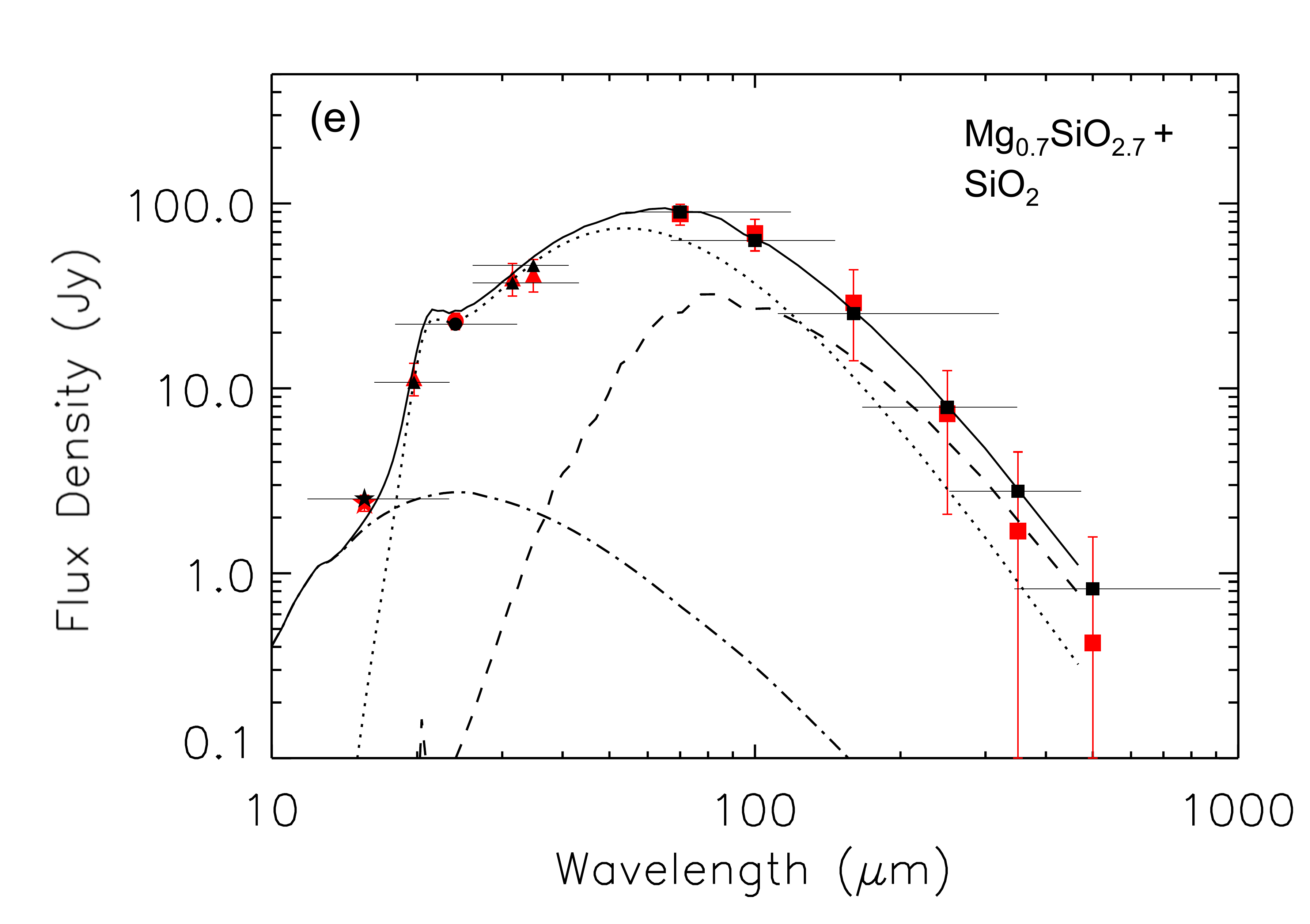}
\epsscale{0.37} \plotone{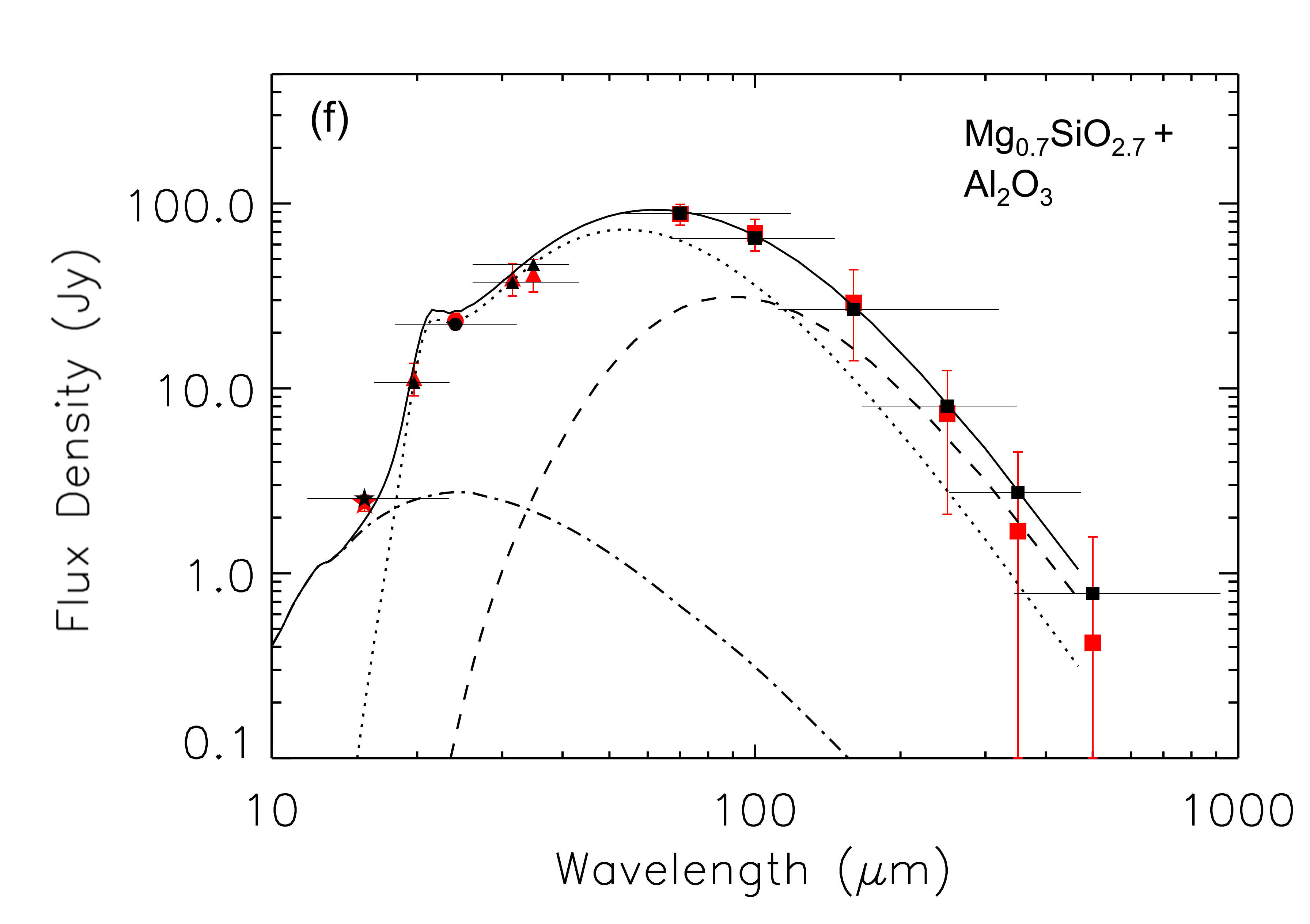}
\caption{\label{dustfits}The total SED of the G54.1+0.3 shell, including the \textit{AKARI}, SOFIA, \textit{Spitzer}, and \textit{Herschel} data points as the filled red stars, triangles, circles, and squares, respectively (listed in Table \ref{tab1}). The fits in all panels include a carbon dust component emitting at a temperature of 150~K (dashed-dotted line), as found from the best fit to the IRS spectra in top panels of Figure~\ref{irsspec}. The additional component included in panel (a) is a single temperature $\rm Mg_{0.7}SiO_{2.7}$ dust composition from \citet{jager03}. The SEDs in the remaining panels are fitted by two components in addition to the hot carbon dust. The dotted curve in all panels represents the $\rm Mg_{0.7}SiO_{2.7}$ composition while the dashed curves in various panels represent (b) $\rm Mg_{0.7}SiO_{2.7}$, (c) amorphous carbon from \citet{rouleau91}, (d) silicates from \citet{weingartner01}, (e) $\rm SiO_2$ from \citet{henning97}, and (f) $\rm Al_2O_3$ from \citet{begemann97}. The black data points represent the model spectra convolved with the filter profiles for each band, with the horizontal black lines represent the total width of the bandpass. In order to fit the observed fluxes, we applied extinction to the model's filter-integrated data points, which explains the slight mismatch between these points and the unabsorbed spectral model. The corresponding dust temperature and masses are listed in Table \ref{tab2}.}
\end{figure*}

\section{ANALYSIS OF THE DUST EMISSION}\label{analysis}

\subsection{Extraction of Source Fluxes}

The source fluxes for the entire IR shell in G54.1+0.3 were measured from the \textit{AKARI}, SOFIA, \spitzer MIPS 24 \micron, and \herschel PACS and SPIRE images that are shown in Figure~\ref{imgpanel}. The SOFIA FORCAST fluxes were extracted from the images using a circular aperture with a radius of 1\farcm0, centered on the IR shell. Since the observations were performed using the chop/nod technique, the background was already removed from the images. The \textit{AKARI}, \spitzer and \herschel fluxes were extracted using an aperture around the IR shell, approximately a circle with a $\sim$ 1\farcm5 radius, and a 4\farcm1 $\times$ 4\farcm5 rectangular region for the background, centered on the shell and excluding the source extraction region. We used the average value in the rectangular region as the background flux and the standard deviation of the pixel-to-pixel fluxes within the region as the background uncertainty. Significant spatial variations in the brightness of the background are evident in the 160-500~\micron\ images shown in Figure~\ref{imgpanel}. In calculating the average background in the SPIRE 250-500~\micron\ images, we masked the brightest cores of the surrounding clouds. The overall uncertainties in the measured flux densities longward of 100 \micron\ are dominated by background confusion. The foreground/background stellar sources that appear in the 15 \micron\ \textit{AKARI} image, but not at 24 \micron, were excluded from the flux measurement.

The IR flux densities for the G54.1+0.3 shell are listed in Table~\ref{tab1}. We applied an extinction correction to the 15.0--34.8~\micron\ flux densities using the mid-IR interstellar extinction law derived by \citet{xue16} from the SDSS- III/APOGEE spectroscopic survey. For the bands that were not explicitly calculated in \citet{xue16}, we used the average of the \citet{wang15} and the \citet{weingartner01} $R_{\rm V}$=5.5 extinction curves \citep[see Figure~18 of][]{xue16}. These two curves diverge only for our 15 and 19 \micron\ data points. In deriving the corrections, we assume $A_{\rm V}$=7.3 \citep{kim13}, leading to $A_{\rm K}$=0.82. The final extinction correction factors that were used in the SED fitting are listed in Table~\ref{tab1}.

\begin{figure*}
\center
\epsscale{0.37} \plotone{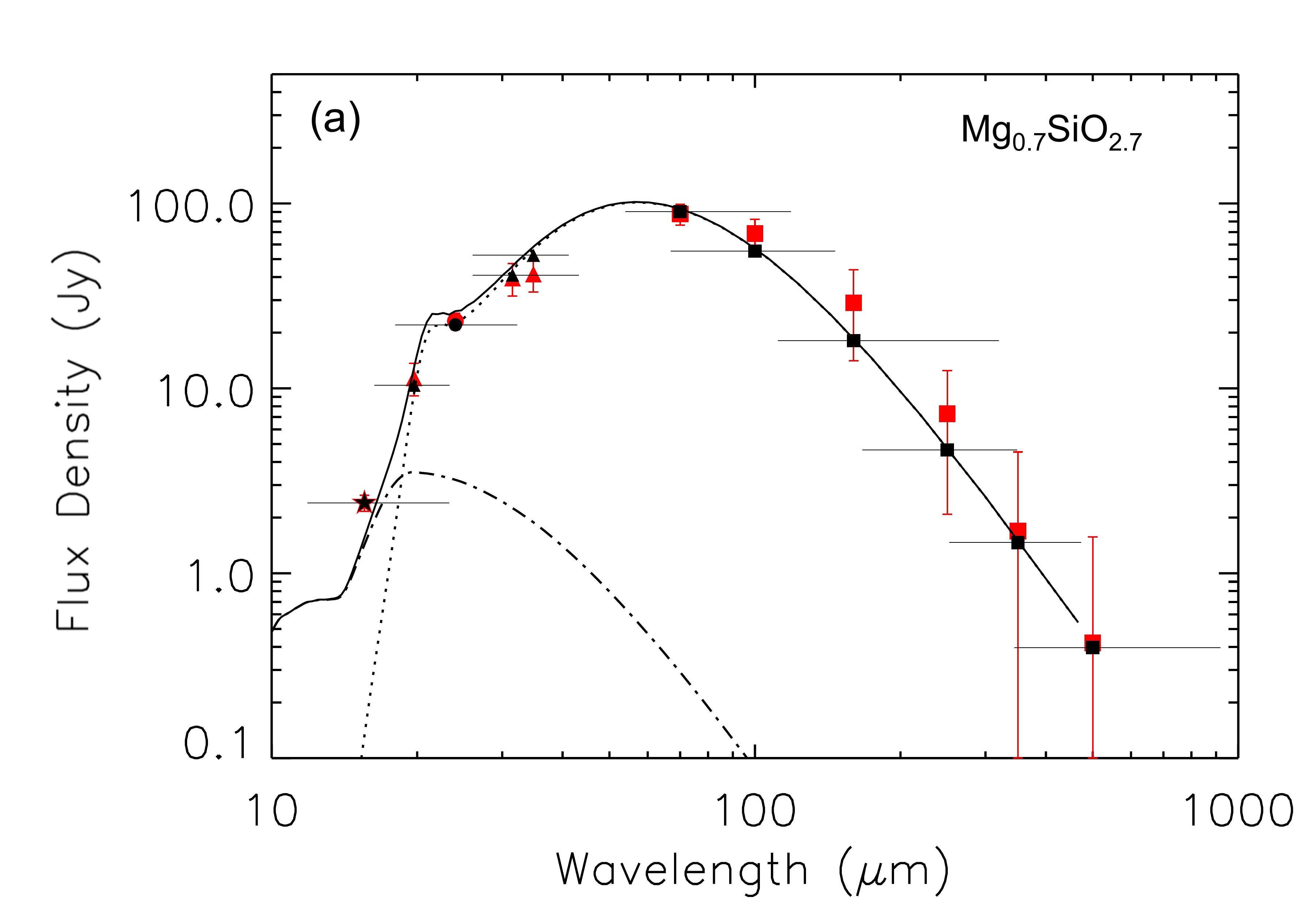}
\epsscale{0.37} \plotone{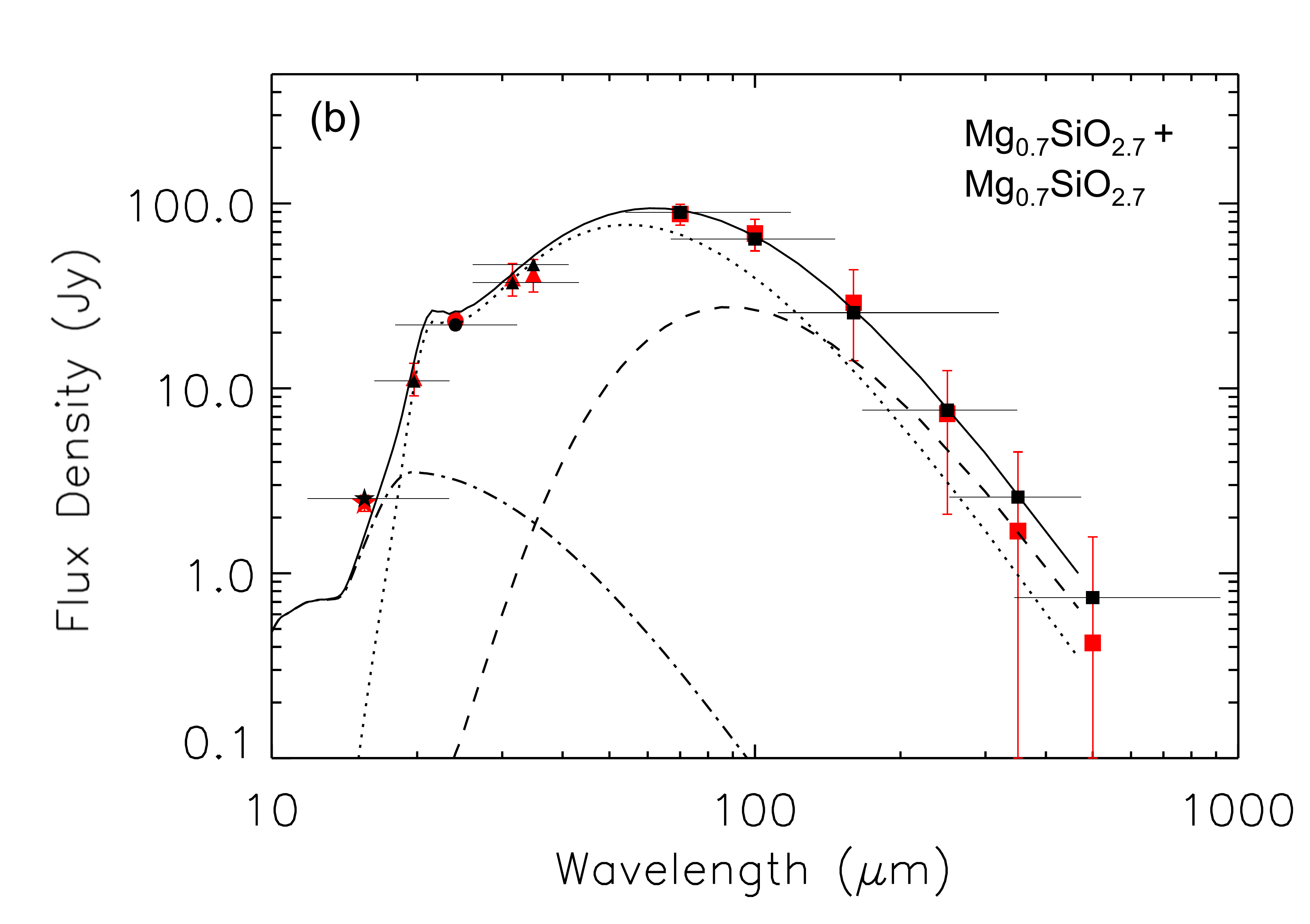}
\epsscale{0.37} \plotone{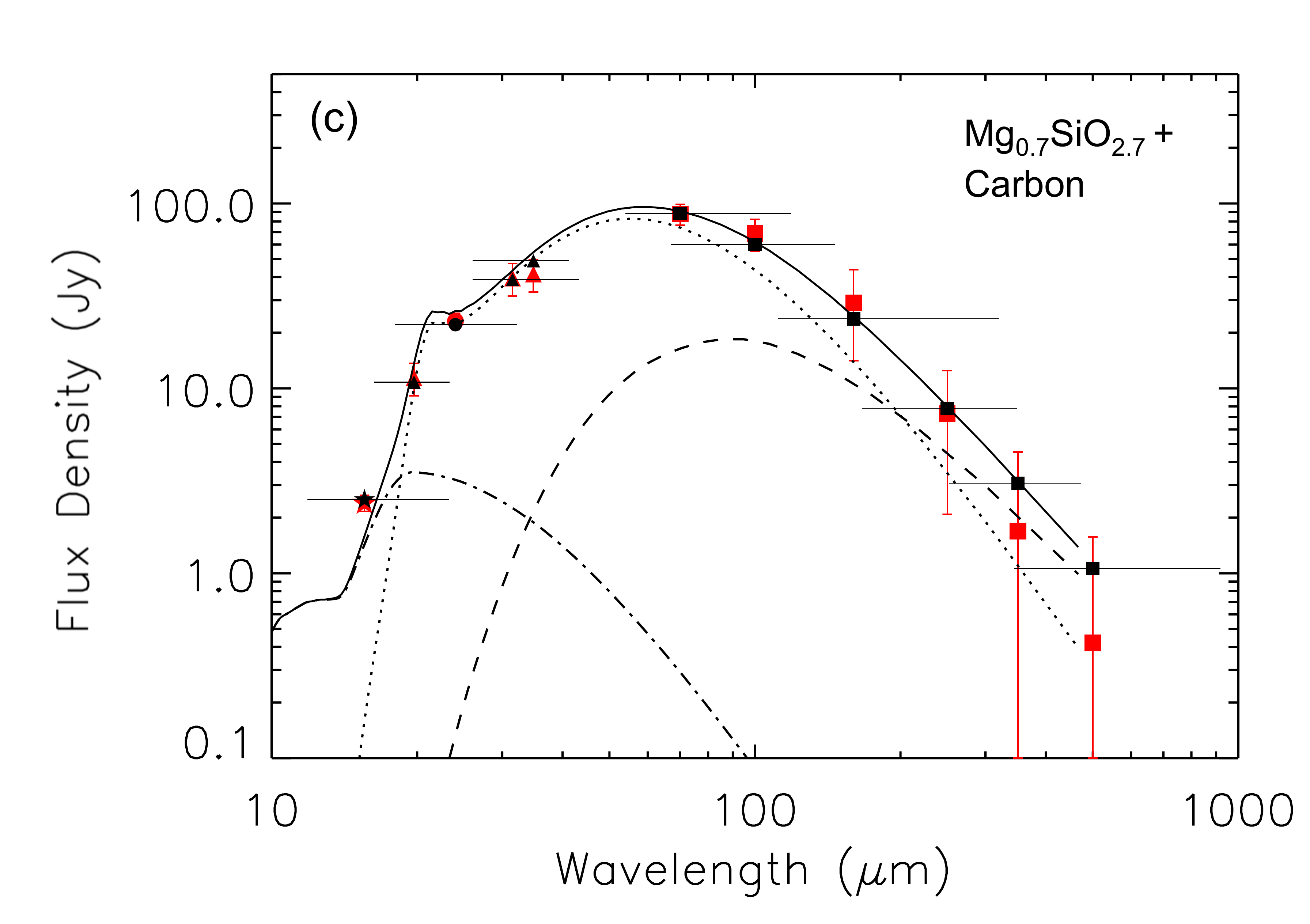}
\epsscale{0.37} \plotone{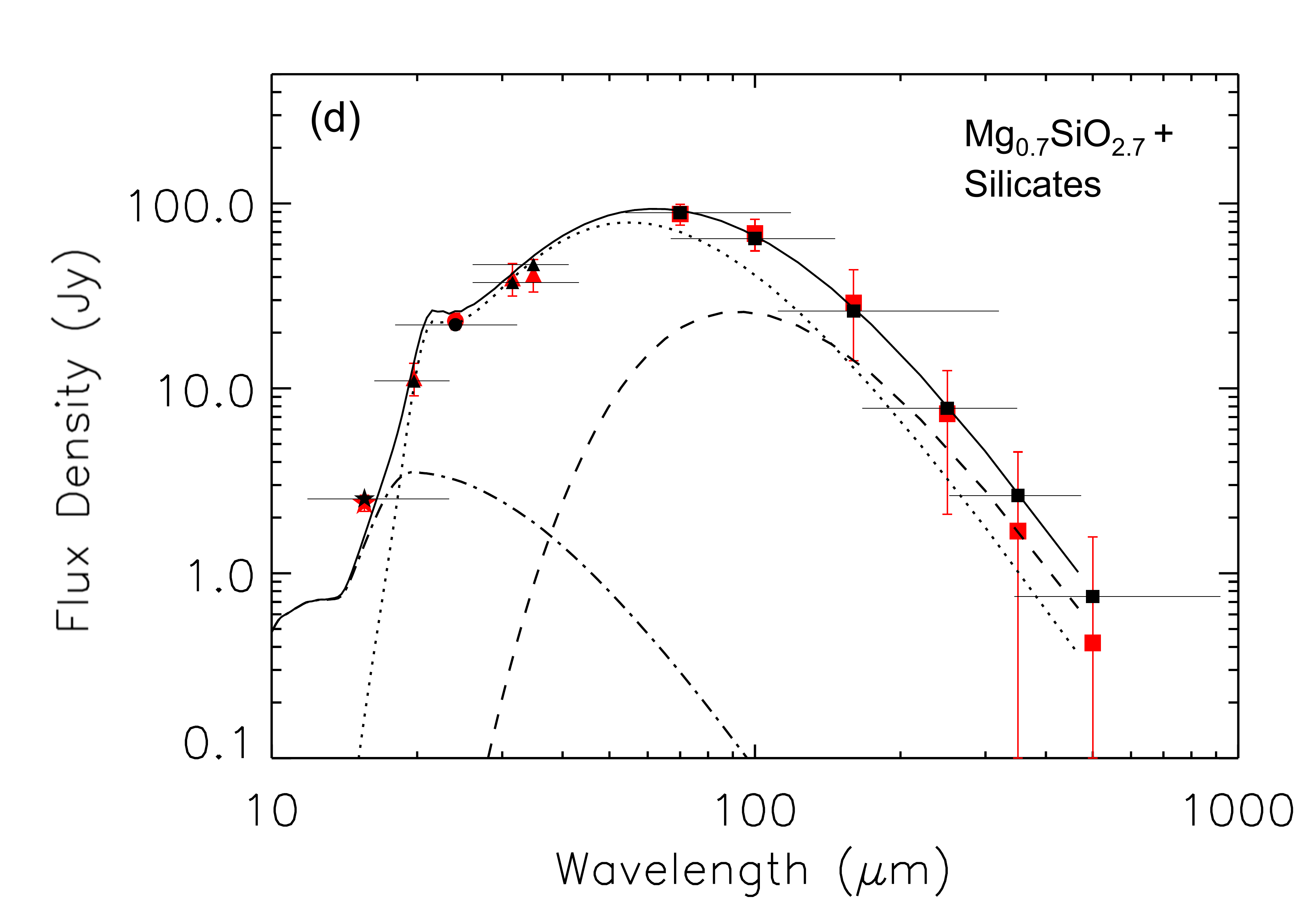}
\epsscale{0.37} \plotone{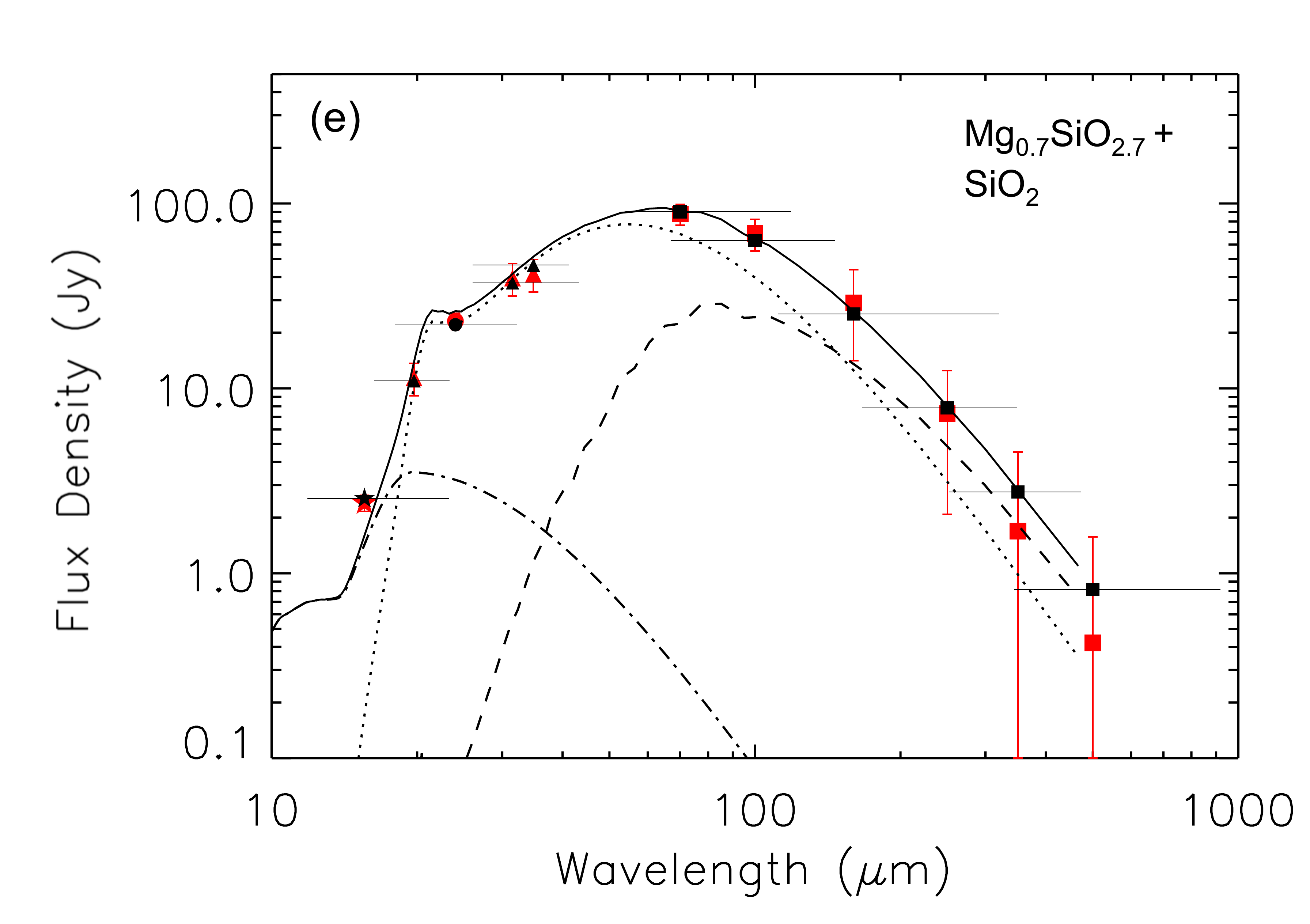}
\epsscale{0.37} \plotone{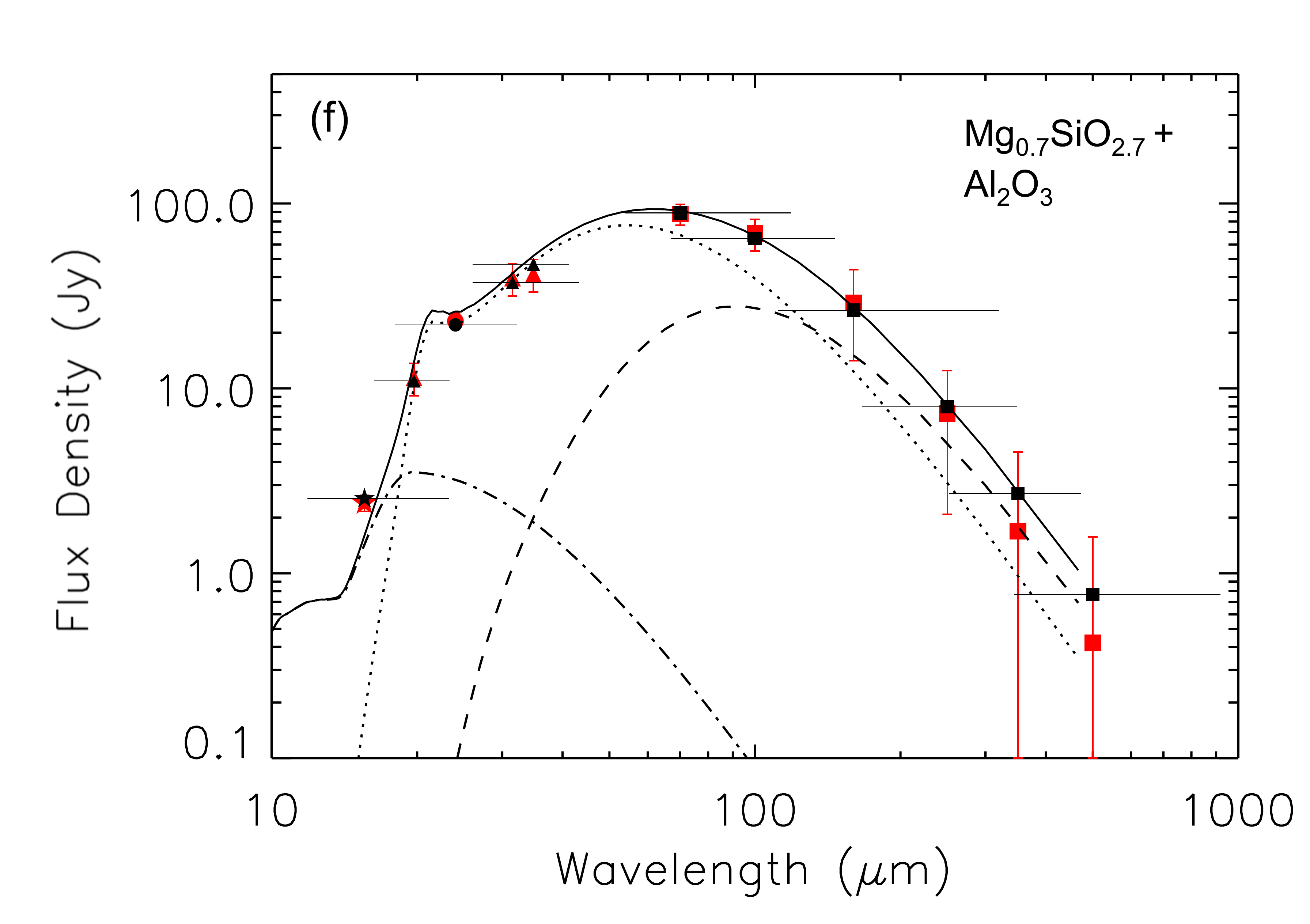}
\caption{\label{dustfits2}The same as Figure~\ref{dustfits}, except that here, the hot component is composed of silicate grains emitting at a temperature of $\sim$~140~K, as found from the fits to the IRS spectra shown in the bottom panels of Figure~\ref{irsspec}.}
\end{figure*}

\subsection{Fitting of the IRS Spectra}\label{specfit}

In Figure~\ref{irsspec}, we show the line-subtracted continuum emission at the two IRS positions shown in Figure~\ref{3color}. The prominent emission feature at $\sim$ 21 \micron\ has been attributed to $\rm Mg_{0.7}SiO_{2.7}$ dust grains \citep{jager03} and used by \citet{arendt14} to fit the dust continuum in Cas A. The mass absorption coefficient that clearly shows this feature is plotted as the purple curve in Figure~\ref{kappas}. The spectra in Figure~\ref{irsspec} show that the feature is much more prominent in position 2 than position 1. We find that the IRS spectra are well fitted by a hotter dust component of either carbon or silicate grains, plus a $\rm Mg_{0.7}SiO_{2.7}$ dust composition emitting at cooler temperature. The hot component has a best-fit temperature of 147 $\pm$ 2 K (152 $\pm$ 1 K) for carbon grains and 146 $\pm$ 1 (133 $\pm$ 1) for silicate grains at positions 1 (2). The best-fit temperatures of the $\rm Mg_{0.7}SiO_{2.7}$ dust are 47 $\pm$ 1 and 58 $\pm$ 1 at positions 1 and 2, respectively, independent of whether the carbon or silicate grains are used to model the hot dust component.
The fits indicate that the hot dust component dominates the spectra shortward of 17~\micron, while the warm component dominates at longer wavelengths. The difference in the shapes of the mid-IR continua at the two positions appear to be caused by variations in the dust temperature of the warm component and not by variations in the grain composition, with the less prominent 21~\micron\ feature at position~1 being an effect of a colder dust temperature that peaks at longer wavelengths.

\begin{deluxetable*}{llcccccc}
\small
\tablecolumns{8} \tablewidth{0pc} \tablecaption{\label{tab2}DUST PARAMETERS}
\tablehead{
\colhead{Comp 1} & \colhead{Comp 2} &  \colhead{$\chi^2$/dof} & \colhead{$\rm T_{1}$ (K)} & \colhead{$\rm T_{2}$ (K)} & \colhead{M$\rm_{1}$ ($\rm M_{\odot}$)} &  \colhead{M$\rm_{2}$ ($\rm M_{\odot}$)} & \colhead{$\rm M_{tot}\: d^2_{6.0}$ ($\rm M_{\odot}$)}
}
\startdata
\cutinhead{Hot dust component: Carbon}
\\
\sidehead{Including long-wavelength residual emission:}
\\
Mg$_{0.7}$SiO$_{2.7}$ & \nodata & 4.6 & 47  $\pm$ 5 & \nodata & 1.74 $\pm$ 0.38 & \nodata & 1.74 $\pm$ 0.38
\\
\\
Mg$_{0.7}$SiO$_{2.7}$ & Mg$_{0.7}$SiO$_{2.7}$ & 1.3 & 49 $\pm$ 2  & 32 $\pm$ 9 & 1.00 $\pm$ 0.57 & 4.31 $\pm$ 4.19 & 5.32 $\pm$ 4.44
\\
\\
Mg$_{0.7}$SiO$_{2.7}$ & Carbon & 2.5 & 49 $\pm$ 2 & 44 $\pm$ 28 & 1.16 $\pm$ 0.88 & 0.046 $\pm$ 0.046 & 1.21 $\pm$ 0.92 
\\
\\
Mg$_{0.7}$SiO$_{2.7}$ & Silicates  & 1.2 & 49 $\pm$ 2 & 32 $\pm$ 8 & 1.03 $\pm$ 0.50 & 0.38 $\pm$ 0.38 & 1.41 $\pm$ 0.70
\\
\\
Mg$_{0.7}$SiO$_{2.7}$ & SiO$_2$  & 1.5 & 49 $\pm$ 2 & 33 $\pm$ 8 & 1.02 $\pm$ 0.51 & 1.60 $\pm$ 1.56 &  2.63 $\pm$ 1.76
\\
\\
Mg$_{0.7}$SiO$_{2.7}$ & Al$_2$O$_3$  & 1.4 & 49 $\pm$ 2 & 38 $\pm$ 14 & 1.00 $\pm$ 0.66 & 0.079 $\pm$ 0.079 & 1.08 $\pm$ 0.70
\\
\\
\sidehead{Excluding long-wavelength residual emission:}
\\
Mg$_{0.7}$SiO$_{2.7}$ & \nodata & 2.0 & 48  $\pm$ 1 & \nodata & 1.52 $\pm$ 0.20 & \nodata & 1.52 $\pm$ 0.20
\\
\\
Mg$_{0.7}$SiO$_{2.7}$ & Mg$_{0.7}$SiO$_{2.7}$ & 1.4 & 49 $\pm$ 3  & 40 $\pm$ 32 & 0.89 $\pm$ 2.75 & 1.23 $\pm$ 1.48 & 2.13 $\pm$ 1.78
\\
\\
Mg$_{0.7}$SiO$_{2.7}$ & Carbon & 1.79 & 48 $\pm$ 1 & 44 $\pm$ 82 & 1.33 $\pm$ 0.96 & 0.015 $\pm$ 0.049 & 1.34 $\pm$ 0.99
\\
\\
Mg$_{0.7}$SiO$_{2.7}$ & Silicates  & 1.4 & 49 $\pm$ 1 & 39 $\pm$ 14 & 0.99 $\pm$ 0.65 & 0.10 $\pm$ 0.11 & 1.09 $\pm$ 0.69
\\
\\
Mg$_{0.7}$SiO$_{2.7}$ & SiO$_2$  & 0.94 & 49 $\pm$ 2 & 46 $\pm$ 4 & 0.90 $\pm$ 0.48 & 0.33 $\pm$ 0.33 &  1.22 $\pm$ 0.22
\\
\\
Mg$_{0.7}$SiO$_{2.7}$ & Al$_2$O$_3$  & 1.6 & 49 $\pm$ 2 & 43 $\pm$ 29 & 1.12 $\pm$ 0.87 & 0.03 $\pm$ 0.03 & 1.14 $\pm$ 0.88
\\

\cutinhead{Hot dust component: Silicates}
\\
\sidehead{Including long-wavelength residual emission:}
\\
Mg$_{0.7}$SiO$_{2.7}$ & \nodata & 4.1 & 47  $\pm$ 1 & \nodata & 1.80 $\pm$ 0.25 & \nodata & 1.80 $\pm$ 0.25
\\
\\
Mg$_{0.7}$SiO$_{2.7}$ & Mg$_{0.7}$SiO$_{2.7}$ & 1.4 & 49 $\pm$ 2  & 31 $\pm$ 9 & 1.13 $\pm$ 0.62 & 4.21 $\pm$ 4.21 & 5.34 $\pm$ 4.71
\\
\\
Mg$_{0.7}$SiO$_{2.7}$ & Carbon & 2.4 & 48 $\pm$ 1 & 43 $\pm$ 28 & 1.32 $\pm$ 0.81 & 0.045 $\pm$ 0.045 & 1.36 $\pm$ 0.85
\\
\\
Mg$_{0.7}$SiO$_{2.7}$ & Silicates  & 1.3 & 48 $\pm$ 2 & 31 $\pm$ 9 & 1.14 $\pm$ 0.55 & 0.38 $\pm$ 0.38 & 1.52 $\pm$ 0.76
\\
\\
Mg$_{0.7}$SiO$_{2.7}$ & SiO$_2$  & 1.5 & 48 $\pm$ 2 & 32 $\pm$ 9 & 1.14 $\pm$ 0.56 & 1.57 $\pm$ 1.57 &  2.71 $\pm$ 1.86
\\
\\
Mg$_{0.7}$SiO$_{2.7}$ & Al$_2$O$_3$  & 1.4 & 49 $\pm$ 2 & 37 $\pm$ 15 & 1.13 $\pm$ 0.71 & 0.077 $\pm$ 0.077 & 1.20 $\pm$ 0.76
\\
\\
\sidehead{Excluding long-wavelength residual emission:}
\\
Mg$_{0.7}$SiO$_{2.7}$ &  \nodata & 1.7  & 47  $\pm$ 1 & \nodata & 1.55 $\pm$ 0.20 & \nodata & 1.55 $\pm$ 0.20
\\
\\
Mg$_{0.7}$SiO$_{2.7}$ & Mg$_{0.7}$SiO$_{2.7}$ & 1.2 & 50 $\pm$ 6  & 42 $\pm$ 43 & 0.66 $\pm$ 6.24 & 1.31 $\pm$ 5.10 & 1.98 $\pm$ 1.32
\\
\\
Mg$_{0.7}$SiO$_{2.7}$ & Carbon & 1.5 & 48 $\pm$ 1 & 45 $\pm$ 91 & 1.37 $\pm$ 1.02 & 0.014 $\pm$ 0.048 & 1.38 $\pm$ 1.06
\\
\\
Mg$_{0.7}$SiO$_{2.7}$ & Silicates  & 1.1 & 49 $\pm$ 1 & 39 $\pm$ 14 & 1.01 $\pm$ 0.66 & 0.11 $\pm$ 0.11 & 1.12 $\pm$ 0.70
\\
\\
Mg$_{0.7}$SiO$_{2.7}$ & SiO$_2$  & 0.8 & 49 $\pm$ 1 & 45 $\pm$ 5 & 0.95 $\pm$ 0.44 & 0.34 $\pm$ 0.34 &  1.29 $\pm$ 0.26
\\
\\
Mg$_{0.7}$SiO$_{2.7}$ & Al$_2$O$_3$  & 1.3 & 49 $\pm$ 1 & 41 $\pm$ 37 & 1.23 $\pm$ 0.86 & 0.024 $\pm$ 0.050 & 1.26 $\pm$ 0.88
\enddata
\tablecomments{Temperatures and masses for the individual warm and cold dust components fits to the global SED. The contributions of the hot carbon (top half of table) and silicate (bottom half of table) components were kept constant for all fits. The dust masses were calculated based on an assumed distance of 6.0 kpc. The table also lists the best fit model parameters for the global SED fits after the subtraction of the long wavelength residual emission that may arise from the background (see Section~\ref{maps}).}
\end{deluxetable*}

%

%
%

\subsection{Global SED Fits}\label{globalfit}

The global SED of the IR shell is shown in Figures~\ref{dustfits} and \ref{dustfits2}, with individual flux density values listed in Table~\ref{tab1}. \citet{temim10} show that the 21 \micron\ spectral feature is present in all regions covered by the \spitzer low-resolution IRS slits, and since the slits span the entire length of the IR shell in the east/west direction, this suggests that the feature is likely present throughout the shell. We also verified that the hot dust component arising from the fits to the IRS high-resolution spectra in Figure~\ref{irsspec} is also present in all regions covered by the low-resolution slits.
We therefore included the hot dust component from either carbon grains (Figure~\ref{dustfits}) or silicate grains (Figure~\ref{dustfits2}), as well as a contribution from $\rm Mg_{0.7}SiO_{2.7}$ grains in all of our SED fits. While an additional third component was not statistically significant in the fit, we explored the possibility of its presence using various grain composition.

As part of our fitting method, we convolved the model spectra with the spectral response of each instrument to derive the expected flux density in each bandpass, applied extinction to each of the values by dividing by the extinction factors in Table~\ref{tab1}, and then fitted the resulting values to the observed flux densities. Figures~\ref{dustfits} and \ref{dustfits2} show the observed flux densities and their uncertainties in red, the best-fit model spectra as the black curves, and the bandpass-integrated and reddened model flux densities in black, with the filter widths indicated by the horizontal black lines. 

We first fitted the global SED with two dust components; a hot dust component of either carbon or silicate grains, and a cooler component of $\rm Mg_{0.7}SiO_{2.7}$ grains. We fixed the hot component's temperatures to those found in the spectral fits in Figure~\ref{irsspec}, 150 K for carbon and 140 K for silicates, while letting their normalizations vary. The resulting best-fit normalization leads to a dust mass of (4.6 $\pm$ 1.2) $\times \: 10^{-5}$ $\rm M_{\odot}$ and (2.2 $\pm$ 0.5) $\times \: 10^{-5}$ $\rm M_{\odot}$ for carbon and silicate grains, respectively. The resulting relative normalizations of the hot and warm components are consistent with those found for the spectral fits.

We then performed fits to the global SED using an additional (third) cold component composed of various grain compositions. In addition to trying an additional cold component of $\rm Mg_{0.7}SiO_{2.7}$ grains (Figure~\ref{dustfits}b), we also tested amorphous carbon \citep{rouleau91},  silicates \citep{weingartner01}, $\rm SiO_2$ \citep{henning97}, and $\rm Al_2O_3$  \citep{begemann97} grains. The mass absorption coefficients for these grain species are shown in Figure~\ref{kappas}, while the best fits to the SED are shown in Figure~\ref{dustfits}c--\ref{dustfits}f with carbon as the hot component, and Figure~\ref{dustfits2}c--\ref{dustfits2}f with silicates as the hot component. The best-fit dust temperatures and masses for all models in Figures~\ref{dustfits} and \ref{dustfits2} are listed in Table~\ref{tab2} and will be discussed in detail in Sections~\ref{temp}~and~\ref{mass}.

The hot carbon/silicate plus a single warm $\rm Mg_{0.7}SiO_{2.7}$ model gives a best-fit temperature of 47 K and a mass of 1.8 $\pm$ 0.3 $\rm M_{\odot}$, assuming a distance of 6.0 kpc. The implausibly large dust mass suggests that the $\rm Mg_{0.7}SiO_{2.7}$ grains are not the sole contributors to the mid-IR SED and that other grain compositions are likely also present.
The addition of another composition resulted in a lower $\chi^2$ for all models, with carbon grains providing a somewhat higher $\chi^2$ than the others due to its flatter slope in emissivity at longer wavelengths (see Figure~\ref{kappas}). The best-fit temperature of the primary $\rm Mg_{0.7}SiO_{2.7}$ dust component is $\sim$ 47--49~K for all model combinations. The temperature and mass of the secondary dust composition are not well constrained and this produces very large uncertainties in the relative masses of the different grain compositions as well as in the total dust mass in the shell.

The relative flux densities of the best fit models plotted in Figures~\ref{dustfits} and \ref{dustfits2} appear to be consistent with the observed morphologies in Figure~\ref{imgpanel}. For instance, as mentioned in Section~\ref{morph}, the morphology at 100 \micron\ and below is different than the observed structure at 250~\micron. The best fits to the global SED show that the emission up to 100 \micron\ is dominated by the warm dust component, while the colder dust component that may have a different spatial morphology dominates at 250 \micron\ and above. In the 160 \micron\ band, the warm and cold component have comparable contributions. In later sections, we will explore the possibility that the emission at 250 \micron\ and above actually arises from a background cloud.

\subsection{Spatially Resolved Fits: Dust Mass \& Temperature Maps}\label{maps}

\begin{figure*}
\epsscale{0.55} \plotone{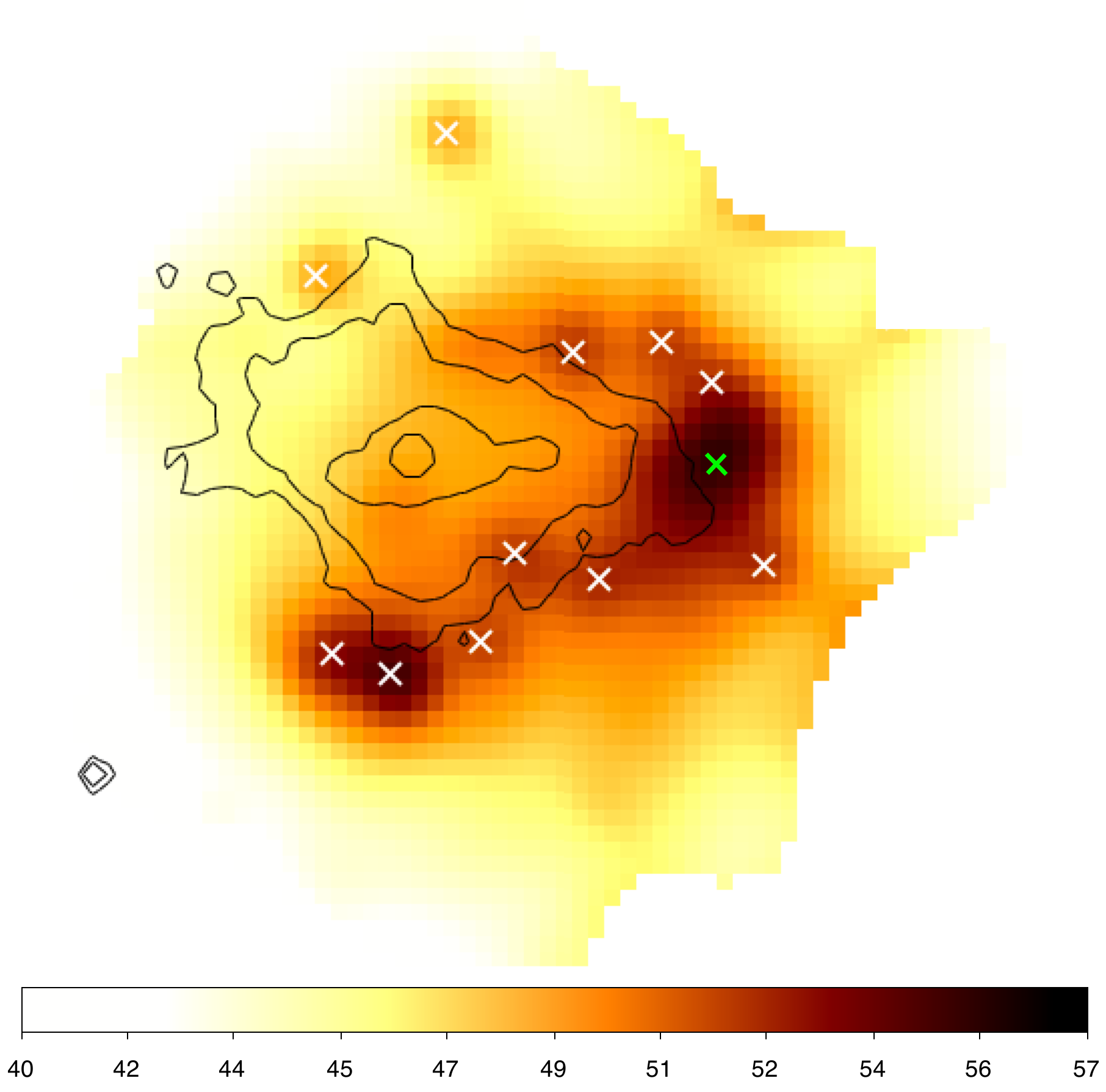}
\epsscale{0.6} \plotone{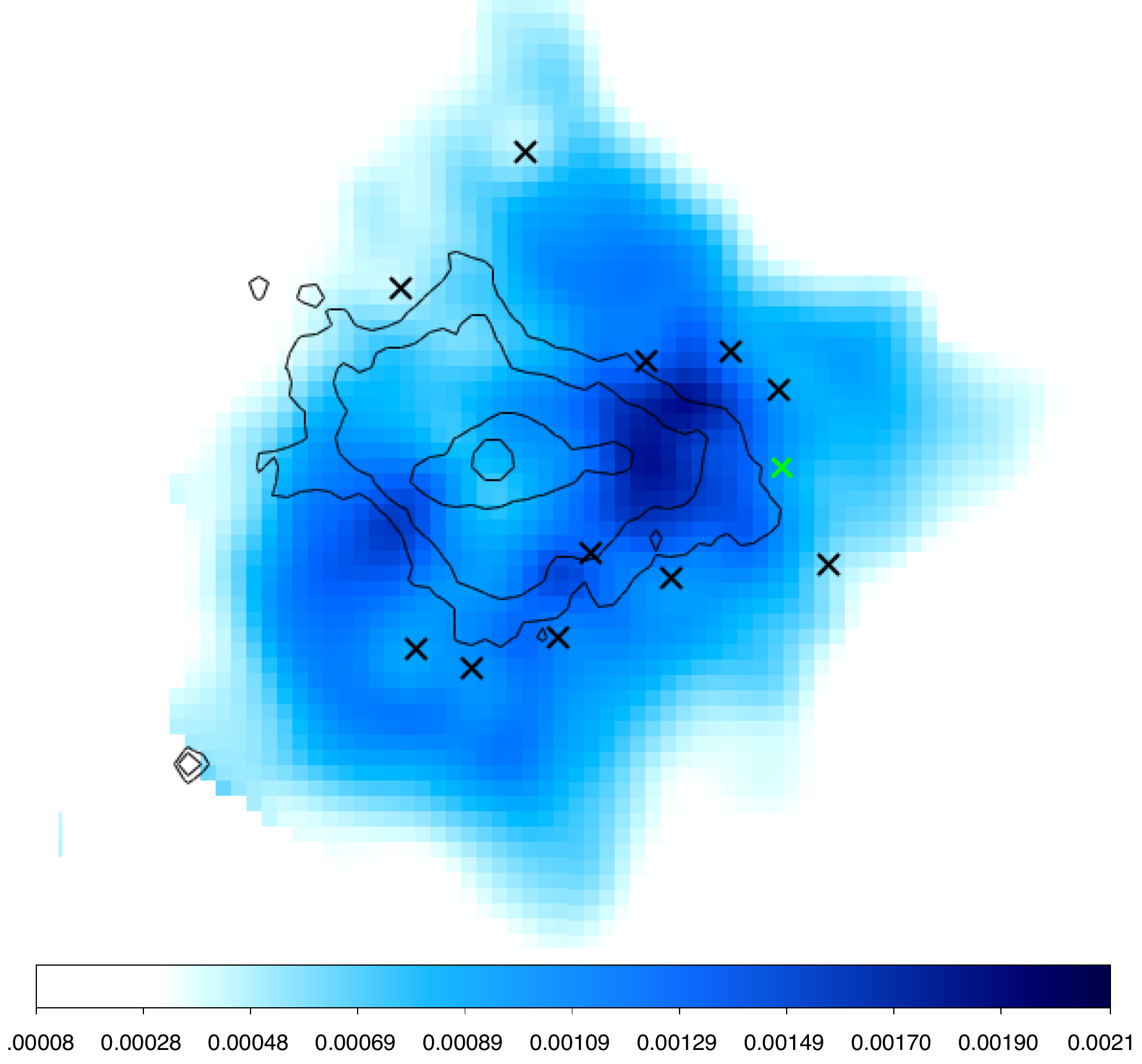}
\caption{\label{dustmaps} Temperature map in units of Kelvin (left) and the dust mass surface density map in units of $\rm M_{\odot}$/pixel, where the pixel size is 3\farcs2 (right), resulting from the fits to the individual pixels' 15-100 \micron\ SEDs, using the $\rm Mg_{0.7}SiO_{2.7}$ dust composition. The ``x" symbols represent the location of the O and B stars (those identified by \citet{koo08} in white and \citet{morris10} in green), while the black contours represent the X-ray PWN. The temperature map clearly shows that the dust temperature peaks around the stars, implying that these O and B stars are the primary heating sources for the dust. The distribution of the dust mass surface density does not show correlation with the stellar sources, but instead reveals a region of enhanced density at the western boundary of the PWN.}
\vspace{10mm}
\end{figure*}

In order to explore the spatial variations in the dust temperature and mass of the shell, we fitted the SEDs of individual pixels across the shell using a single dust component of $\rm Mg_{0.7}SiO_{2.7}$ grains. In order to preserve the spatial resolution, minimize the effects of the background confusion, and also to only model wavelengths that appear to have similar morphologies, we only used the 15--100~\micron\ images in our fits. The SOFIA images were also excluded due to the poor sensitivity to the fainter emission in the shell. The images were all convolved to the PACS 160 \micron\ resolution using the \citet{aniano11} convolution kernels, resulting in a pixel size of 3\farcs2. The SED fits provided a temperature and corresponding dust mass for each pixel. These maps are shown in Figure~\ref{dustmaps}, with the left panel showing the temperature map and the right panel showing the dust mass surface density map. The black contours represent the \chandra X-ray emission from the PWN seen in Figure~\ref{3color}, while the ``x" symbols represent the locations of the stars identified by \citet{koo08} and \citet{morris10}.

The temperature in the shell varies from 42~--~57 K with an average temperature and standard deviation of 46 $\pm$ 4 K. The sum of the individual pixels in the dust mass surface density map leads to a total dust mass of $\rm 1.8 \pm 0.3 \:M_{\odot}$, consistent with the mass derived from fitting the integrated SED of the shell using only $\rm Mg_{0.7}SiO_{2.7}$ grains (see Table~\ref{tab2}). The dust mass and temperature maps are consistent with the physical scenario proposed by \citet{temim10} in which the stellar sources heat the surrounding SN-condensed dust, since the temperature clearly peaks at the locations of the stellar sources, while the dust mass surface density shows no such correlation. The dust mass surface density map does show a slight deficiency in dust mass at the locations of the brightest stellar sources. The most likely explanation for this deficiency is that the temperature at the locations of the stars was slightly overestimated due to the inclusion of 15~\micron\ data that has a significant contribution from hot dust.

\begin{figure*}
\center
\epsscale{1.15} \plotone{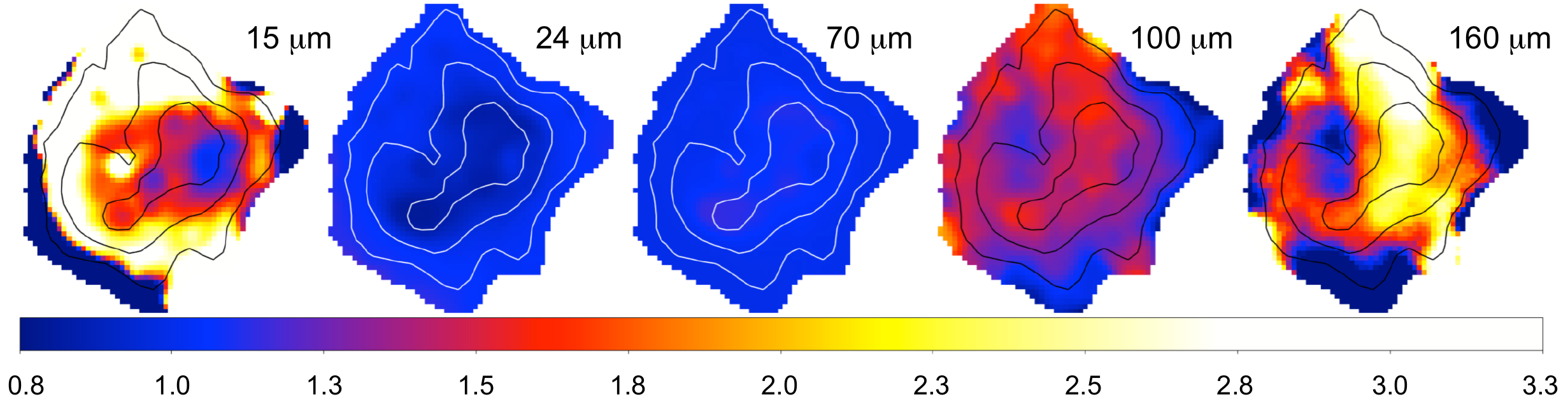}
\caption{\label{ratios}Ratio maps of the background-subtracted AKARI 15 \micron, MIPS 24 \micron, and PACS 70, 100 and 160 \micron\ images and the flux-density maps at corresponding wavelengths derived from the best-fit dust and temperature maps shown in Figure~\ref{dustmaps}. The contours represent the 70~\micron\ shell emission. The ratio of one (blue) indicates that the best-fit spatial model accounts for all of the observed flux, while a higher ratio indicates an observed excess above the best-fit model. The ratio maps provide information about the morphology of the emission excess; the excess at 15 \micron\ has a different morphology than the IR shell observed in the mid-IR, while the excess at 100 and 160 \micron\ resembles the morphology seen at 250 \micron\ and longward, suggesting that it arises from a contribution from a secondary dust component.}
\end{figure*}

In order to determine how well the single-component spatial fits match the global SED, we compared the observed and model-predicted flux densities of the shell at all observed wavebands. While the 19-70 \micron\ data are well represented by the model, the 15 \micron\ flux is underestimated by $\sim$ 35\%. The model also underestimates the observed emission at 100, 160, 250, 350, and 500 \micron, by 28\%, 44\%, 43\%, 23\%, and 16\%, respectively. In order to investigate the nature of the excess emission above the best-fit spatial model, we used the dust mass and temperature maps in Figure~\ref{dustmaps} to produce model-predicted flux density images at 15, 24, 70, 100, and 160 \micron, and then divided the observed images by these maps to produce ratio maps. The ratio maps are shown in Figure~\ref{ratios}, all using the same linear scale (0.8--3.3) and color scheme. The blue color indicates a ratio of $\sim$1, where the model most closely matches the observation, while the values above unity show the spatial distribution of the emission excess that was not well-predicted by the model. The 24 and 70 \micron\ emission is clearly well-described by the model across the entire shell, while there is a clear excess at 15, 100, and 160 \micron. 

The spatial distribution of the excess at 15 \micron\ is particularly interesting. It does not at all correlate with the shell morphology at 24 \micron, but is instead concentrated along a thin shell on the outskirts of the observed mid-IR shell, as well as a couple of point sources, one of which is a background source not associated with the SNR. This suggests that the 15 \micron\ excess emission does not belong to the same dust component that emits at 24 \micron, but instead may be produced by hot dust grains that have a different spatial distribution or by line emission that could partly contribute to the 15 \micron\ \textit{AKARI} band. One hypothesis is that very small dust grains in the shell are heated to temperatures much higher than the equilibrium temperature due to their lower heat capacity, and that their emission is more pronounced at the outskirts of the cluster where the radiation field starts to drop off and where the temperature of the larger grains is cooler. The presence of this hot component is also supported by the fits to the IRS spectra shown in Figure~\ref{irsspec}. We note here that the average contribution of the hot dust component in the IRS fits to the 15~\micron\ AKARI waveband is $\sim$ 35\%, the same fraction that is seen in the 15 \micron\ excess above the best spatial fit (first panel of Figure~\ref{ratios}).

\begin{figure*}
\center
\epsscale{1.2} \plotone{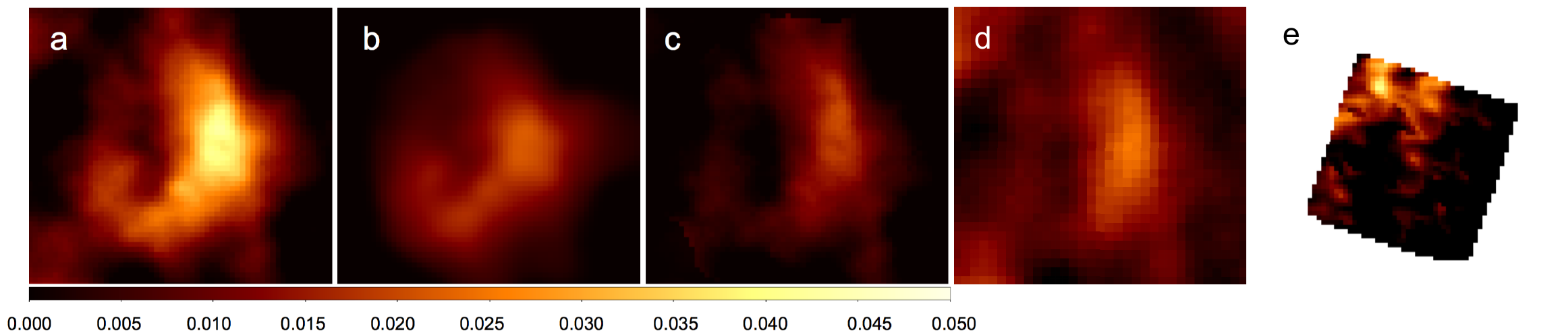}
\caption{\label{160}(a) PACS 160 \micron\ image in units of Jy/pixel. (b) Model-predicted 160 \micron\ flux calculated from the mass and temperature maps in Figure~\ref{dustmaps}. (c) The residual 160 \micron\ emission after the subtraction of the model-predicted flux. Image in panels a--c are on the same brightness scale and stretch. (d) 250 \micron\ SPIRE image that shows the same morphology as the residual 160 \micron\ emission in panel (c), and (e) Map of the [\ion{C}{2}] 157.7 \micron\ line emission that does not spatially correlate with the IR shell, but does show some resemblance to the 15 \micron\ excess emission shown in Figure~\ref{ratios}.}
\end{figure*}

The ratio maps at 100 and 160 \micron\ show that the morphology of the excess emission resembles the morphology in the far-IR. This is seen more clearly in Figure~\ref{160}, where we show the observed 160 \micron\ image in panel (a), the spatial model-predicted 160 \micron\ flux arising from $\rm Mg_{0.7}SiO_{2.7}$ in panel (b), the residual between the observed image and the fitted model in panel (c), and the 250 \micron\ observed image in panel (d). For comparison, we also show a map of the [\ion{C}{2}]~157.7 \micron\ line emission from the PACS spectral observations in Figure~\ref{160}e, which does not seem to correlate with the 160 \micron\ dust emission, but instead shows a spatial distribution more similar to the morphology of the 15 \micron\ excess in Figure~\ref{ratios}. 
The residual map in panel (c) of Figure~\ref{160} clearly shows that the excess emission at 160 \micron\ closely resembles the morphology at 250 \micron, suggesting that the 100 and 160 \micron\ excess above the $\rm Mg_{0.7}SiO_{2.7}$ spatial model either arises from a distinct dust component that dominates the emission at longer wavelengths or from background emission not associated with the SNR. 

\begin{figure}
\center
\epsscale{1.2} \plotone{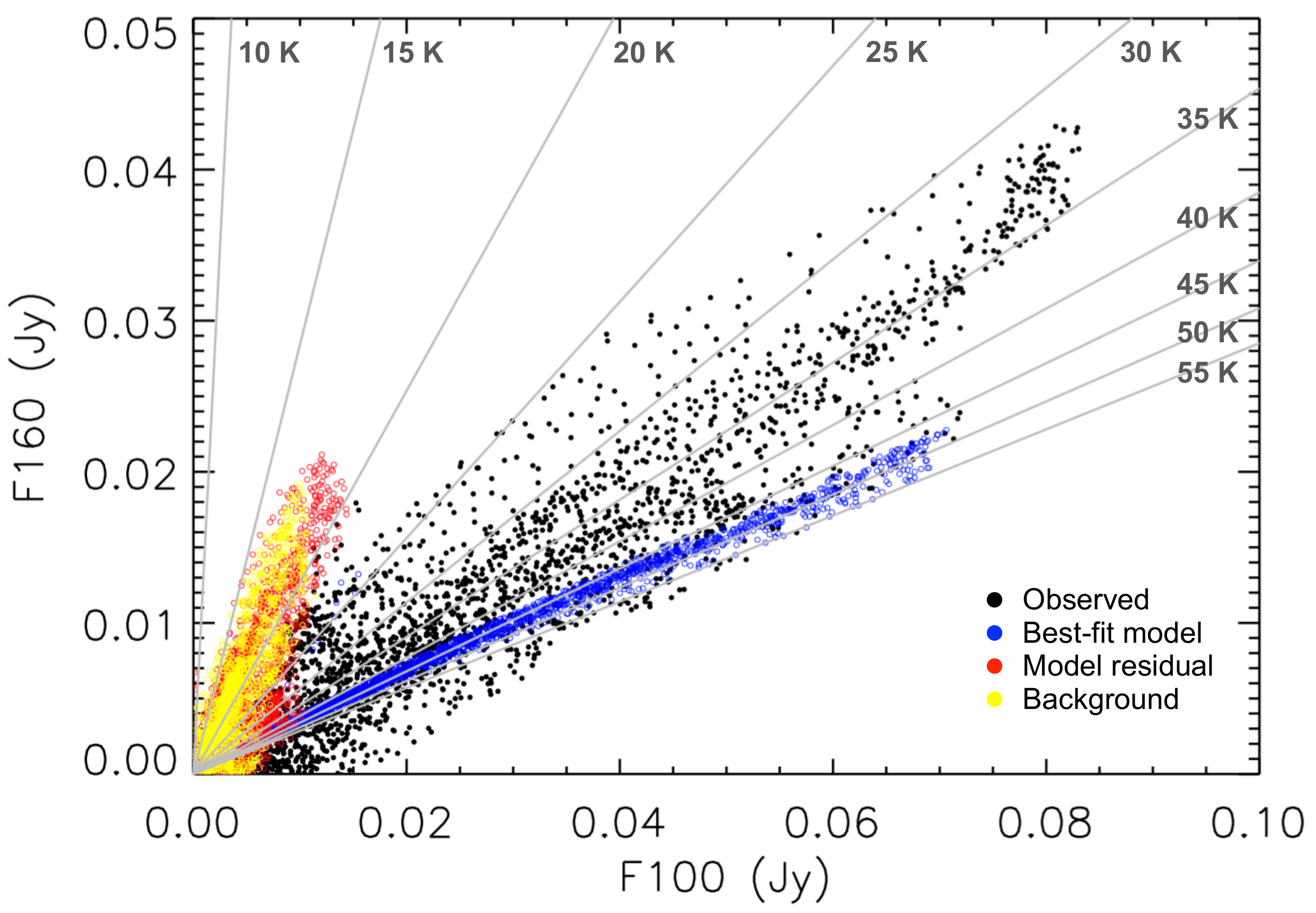}
\caption{\label{colorplot}A scatter plot of the 100 \micron\ flux density versus the 160 \micron\ flux density for the observed emission in the IR shell (black), best-fit spatial model using $\rm Mg_{0.7}SiO_{2.7}$ grains (blue), the observed minus model residual emission (red), and the flux densities from an annular background region around the source (yellow). The solid gray lines represent the 160/100 \micron\ flux density ratios for various temperatures. The best-fit spatial model produces temperature beween 45 and 55 K, while the long-wavelength residual has the same characteristics as the surrounding background.}
\end{figure}

In Section~\ref{globalfit}, we explored the possibility that an additional dust component with a different composition contributes to the global SED of the shell and dominates the emission at wavelengths longer than 100 \micron. The best-fit parameters for the models that include this secondary component are listed in Table~\ref{tab2}. 
Another possibility is that the emission at long wavelengths actually originates from a background cloud, since the morphology at wavelengths longer than 160 \micron\  differs from the morphology at shorter wavelengths. If we assume that the emission at 24-70 \micron\ is dominated by a single dust component of $\rm Mg_{0.7}SiO_{2.7}$, as was assumed in the spatial fit, we can then test whether the level of residual emission at longer wavelengths is consistent with the surrounding background. In Figure~\ref{colorplot}, we show a scatter plot of the 100 versus 160 \micron\ flux densities for the observed emission in the shell, the best-fit spatial model for a single component of $\rm Mg_{0.7}SiO_{2.7}$ grains, the residual emission after the subtraction of the spatial model, and finally, the background emission in an annulus surrounding the source. The gray lines represent the 160/100 \micron\ ratios for various temperature of $\rm Mg_{0.7}SiO_{2.7}$ grains. As can be seen from the plot, the best-fit spatial model is composed of grains emitting in a narrow temperature range between 45 and 55~K (also seen in the left panel of Figure~\ref{dustmaps}), while the long-wavelength residual is seen to have the same properties as surrounding background emission. 

In this scenario, the $\rm Mg_{0.7}SiO_{2.7}$ grains with a temperature and mass distribution shown in Figure~\ref{dustmaps} dominate the emission from 15--100 \micron, while residual background emission significantly contributes to the 160 \micron\ flux and dominates at the far-IR SPIRE wavelengths. We therefore performed additional fits to the global SED, assuming that the residual long-wavelength emission arises from the background. We used the same compositions as in Section~\ref{globalfit}, but subtracted the residual long-wevelength emission from the integrated flux densities. Since the fit, by definition, is well described by a single component of warm $\rm Mg_{0.7}SiO_{2.7}$ grains (in addition to the hot component derived from the spectral fits), a third component is not statistically significant. However, adding this component allowed us to test whether an addition of grains of a different composition affects the best-fit dust parameters. The results of the fits are included in Table~\ref{tab2}, in the subsections indicating that the long-wavelength residual emission has been excluded, and will be discussed in Sections~\ref{temp}~and~\ref{mass}.

\section{Dust Properties}\label{dust}

\subsection{Composition}\label{comp}

The prominent spectral feature that is observed at 21~\micron\ has the same spectral profile as the feature observed in the spectrum of the SN-condensed dust in Cas~A \citep[][]{arendt99, ennis06}. As far as we know, Cas A and G54.1+0.3 are the only two sources whose spectra exhibit a feature of this particular shape. \citet{arendt14} attributed the feature to $\rm Mg_{0.7}SiO_{2.7}$ grains, characterized by an MgO to SiO$_2$ ratio of 0.7. Out of sixty different grain compositions that they tested, this species is the only one that has a 21 \micron\ spectral feature that fits the width and shape of the feature observed in Cas A. As discussed in Section~\ref{specfit}, we confirmed that the same grain species can fit the \spitzer high-resolution spectra of G54.1+0.3 that are shown in Figure~\ref{irsspec}. Since the 21 \micron\ feature is present throughout the low-resolution IRS slits that run across the IR shell \citep{temim10}, we used the $\rm Mg_{0.7}SiO_{2.7}$ grains as the primary component in all our fits, but also included other common grain compositions predicted by dust condensation models in Type~IIP SNe for possible secondary components. For example, the models of \citet{kozasa09} and \citet{sarangi15} predict that 80--100\% of the SN-condensed dust in Type~IIP explosions of $\sim$~20~$\rm M_{\odot}$ progenitors is made up of carbon, alumina, and some form of magnesium silicate grains. The mass absorption coefficients of the grain species used in our fits are shown in Figure~\ref{kappas}, and they include amorphous carbon, silicates, $\rm SiO_2$, and $\rm Al_2O_3$. Another common dust composition predicted by \citet{kozasa09} is MgO, but our fits disqualify a major contribution from this species due its spectral feature at $\sim$~100~\micron\ that does not appear in the SED of the shell.

The best fits in Figure~\ref{dustfits} and Table~\ref{tab2} indicate that if the $\rm Mg_{0.7}SiO_{2.7}$ grains are indeed responsible for producing the feature and account for a significant fraction of the observed emission at mid-IR wavelengths, they are a major component of the total dust mass in the shell. Any additional component is equally well fitted by the other grain species we tested (see Table~\ref{tab2}), with the exception of carbon component, which results in a somewhat larger $\chi^2$. This is due to the fact that the absorption coefficient for carbon grains (see Figure~\ref{kappas}) produces a somewhat flatter slope at longer wavelengths than indicated by the SPIRE data.

\subsection{Temperature}\label{temp}

\begin{figure}
\center
\epsscale{1.2} \plotone{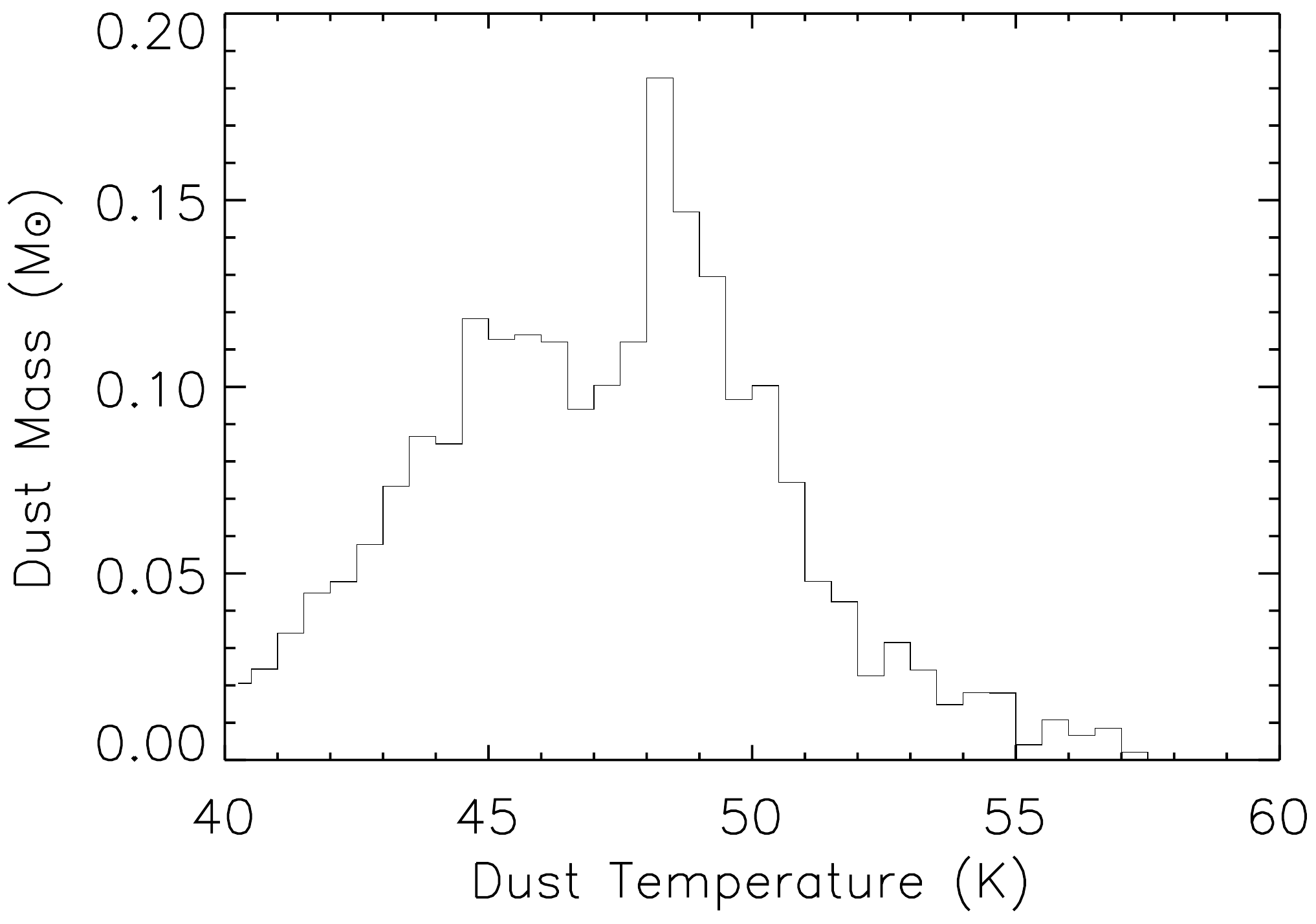}
\caption{\label{dustplot}Distribution of dust mass as function of temperature resulting from the best-fit maps shown in Figure \ref{dustmaps}. The plot shows that most of the dust mass emits in a narrow temperature range between 45 and 51 K. The peak at a temperature of $\sim$ 49 K originates from the enhanced region of surface mass density that is evident in the right panel of Figure~\ref{dustmaps}.}
\end{figure}

The map in Figure~\ref{dustmaps} shows the temperature distribution of the $\rm Mg_{0.7}SiO_{2.7}$ grains. Since the grains at any given location will likely have a distribution of grain sizes, the temperature for any given pixel should be thought of as the average temperature for that location. The temperature of the dust in the shell ranges from 42 to 57~K, with an average temperature and standard deviation of 46 $\pm$ 4 K. The plot in Figure~\ref{dustplot} shows the mass-temperature distribution in the shell, and indicates that most of the mass is actually emitting in a narrow temperature range of 43--52 K. This explains why the mass of the single-temperature warm component fit in Table~\ref{tab2} is similar to that found from the spatially resolved fit, unlike what would be expected for a very wide dust temperature distribution in the shell \citep{temim13}.

The left panel of Figure~\ref{dustmaps} clearly shows that the temperature peaks at the location of the stars that are embedded in the shell, confirming that the stars are the primary heating sources for the surrounding dust. In order to verify that this is plausible, we compared the total luminosity of the eleven O and B stars in the shell analyzed by \citet{kim13} to the total luminosity of the IR SED seen in Figure~\ref{dustfits}. Based on the spectral classification of the stars, we assume a luminosity of 25,000~$\rm L_{\odot}$ for each star. We calculate an IR luminosity of the shell to be $\sim$~9500~$\rm L_{\odot}$, which is only a small fraction of the total luminosity of the stars, approximately 3.5\%. We note that there may be additional stars present inside the volume of the shell for which near-IR spectroscopy was not obtained by these authors. In particular, the bright extended region in the left panel of Figure~\ref{3color} that coincides with position 2 of the IRS slit and also shows up as a hot region in the dust temperature map in Figure~\ref{dustmaps} is likely heated by an early-type star identified by \citet{morris10}, but not studied by \citet{kim13}. The location of this star is marked by the green``x" in Figure~\ref{dustmaps}. This region also has a higher gas density that is likely produced by the compression of the ejecta material by the pulsar's jet \citep{temim10}, but while the jet may contribute to some heating of the dust in this location, the dust temperature distribution and total luminosity imply that the stars are responsible for most of the heating.

The results from the global SED fits in Table~\ref{tab2} show that all model combinations result in $\rm Mg_{0.7}SiO_{2.7}$ grains that are warmer on average than the secondary dust component of a different composition. While a physical dust heating model for the dust in the shell will be explored in a future publication, our preliminary model for heating of dust around a single B0V star \citep[see][]{temim10} confirms that the $\rm Mg_{0.7}SiO_{2.7}$ grains are heated to higher temperatures than carbon or silicate grains with comparable sizes. It furthermore shows that a typical grain size of the $\rm Mg_{0.7}SiO_{2.7}$ dust would need to be relatively large (on the order of $\sim1$~\micron) to achieve the temperatures produced by the best fits in Table~\ref{tab2}.

\begin{deluxetable*}{ccccccccccc}
\tablecolumns{11} \tablewidth{0pc} \tablecaption{Dust Masses Allowed by Nucleosynthetic Yields\label{tab3}}

\tablehead{
\colhead{M$\rm_{*}$ ($\rm M_{\odot}$)} & \multicolumn{5}{c}{$\rm M_i/$ ($\rm M_{\odot}$)} & \colhead{$\rm N_{Mg}/N_{Si}$} & \multicolumn{4}{c}{$\rm M_{d}$ ($\rm M_{\odot}$)} \\
\\
 \cline{2-6}
 \cline{8-11} 
 \\
\colhead{}  & \colhead{C} & \colhead{O} & \colhead{Mg} & \colhead{Al} & \colhead{Si} & \colhead{}  & \colhead{$\rm Mg_{0.7}SiO_{2.7}$} & \colhead{MgO+SiO$_2$} & \colhead{$\rm Al_2O_3$} & \colhead{$\rm Mg_{0.7}SiO_{2.7}+C$}
}
\startdata
\sidehead{Woosley \& Heger (2007):}
13.0	&	0.0949	&	0.5712	&	0.0505	&	0.0041	&	0.0707	&	0.83	&	0.222	&	0.234	&	0.008	&	0.316	\\
14.0	&	0.1154	&	0.7229	&	0.0579	&	0.0046	&	0.0713	&	0.94	&	0.223	&	0.248	&	0.009	&	0.339	\\
15.0	&	0.1336	&	0.8380	&	0.0525	&	0.0041	&	0.0742	&	0.82	&	0.232	&	0.245	&	0.008	&	0.366	\\
16.0	&	0.1577	&	0.9371	&	0.0529	&	0.0041	&	0.0490	&	1.25	&	0.154	&	0.192	&	0.008	&	0.311	\\
17.0	&	0.1768	&	1.3604	&	0.0826	&	0.0064	&	0.1715	&	0.56	&	0.427	&	0.502	&	0.012	&	0.604	\\
18.0	&	0.1960	&	1.6358	&	0.1496	&	0.0137	&	0.1110	&	1.56	&	0.348	&	0.483	&	0.026	&	0.544	\\
19.0	&	0.2077	&	1.8711	&	0.0985	&	0.0093	&	0.1358	&	0.84	&	0.426	&	0.452	&	0.018	&	0.633	\\
20.0	&	0.2336	&	1.9604	&	0.1112	&	0.0089	&	0.2178	&	0.59	&	0.575	&	0.648	&	0.017	&	0.809	\\
21.0	&	0.2701	&	2.5558	&	0.1536	&	0.0146	&	0.1296	&	1.37	&	0.406	&	0.530	&	0.028	&	0.676	\\
22.0	&	0.2591	&	2.5249	&	0.1204	&	0.0117	&	0.2170	&	0.64	&	0.623	&	0.662	&	0.022	&	0.882	\\
23.0	&	0.2986	&	2.5480	&	0.1478	&	0.0119	&	0.2556	&	0.67	&	0.764	&	0.789	&	0.022	&	1.063	\\
24.0	&	0.2825	&	3.0035	&	0.2442	&	0.0207	&	0.2663	&	1.06	&	0.834	&	0.971	&	0.039	&	1.117	\\
25.0	&	0.3629	&	3.3841	&	0.2307	&	0.0204	&	0.2990	&	0.89	&	0.937	&	1.018	&	0.039	&	1.300	\\
26.0	&	0.3331	&	3.7727	&	0.1335	&	0.0154	&	0.3686	&	0.42	&	0.691	&	1.007	&	0.029	&	1.024	\\
27.0	&	0.3159	&	4.0532	&	0.1674	&	0.0198	&	0.2937	&	0.66	&	0.866	&	0.903	&	0.037	&	1.182	\\
28.0	&	0.3793	&	4.1549	&	0.2012	&	0.0229	&	0.1631	&	1.43	&	0.511	&	0.680	&	0.043	&	0.890	\\
29.0	&	0.3828	&	4.8632	&	0.2474	&	0.0285	&	0.1788	&	1.60	&	0.560	&	0.789	&	0.054	&	0.943	\\
30.0	&	0.4324	&	5.1512	&	0.3411	&	0.0328	&	0.1671	&	2.36	&	0.524	&	0.918	&	0.062	&	0.956	\\
40.0	&	0.4501	&	6.4855	&	0.3656	&	0.0396	&	0.1627	&	2.60	&	0.510	&	0.949	&	0.075	&	0.960	\\
60.0	&	6.1495	&	6.5599	&	0.1673	&	0.0168	&	0.2096	&	0.92	&	0.657	&	0.723	&	0.032	&	6.806	\\

\sidehead{Sukhbold et al. (2016):}																							
12.7	&	0.1101	&	0.3886	&	0.0341	&	0.0022	&	0.0481	&	0.81	&	0.151	&	0.159	&	0.004	&	0.261	\\
13.4	&	0.1177	&	0.5650	&	0.0489	&	0.0039	&	0.0551	&	1.03	&	0.172	&	0.198	&	0.007	&	0.290	\\
13.8	&	0.1265	&	0.6189	&	0.0504	&	0.0038	&	0.0600	&	0.97	&	0.188	&	0.211	&	0.007	&	0.314	\\
14.3	&	0.1365	&	0.7046	&	0.0563	&	0.0044	&	0.0606	&	1.08	&	0.190	&	0.222	&	0.008	&	0.326	\\
14.7	&	0.1493	&	0.7608	&	0.0539	&	0.0041	&	0.0615	&	1.01	&	0.192	&	0.220	&	0.008	&	0.342	\\
15.4	&	0.1718	&	0.9740	&	0.0628	&	0.0047	&	0.0825	&	0.89	&	0.259	&	0.279	&	0.009	&	0.430	\\
16.2	&	0.1908	&	1.1489	&	0.0705	&	0.0052	&	0.0994	&	0.81	&	0.312	&	0.328	&	0.010	&	0.503	\\
16.6	&	0.1973	&	1.2113	&	0.0748	&	0.0054	&	0.1047	&	0.82	&	0.329	&	0.347	&	0.010	&	0.526	\\
17.0	&	0.2054	&	1.2862	&	0.0785	&	0.0056	&	0.1102	&	0.82	&	0.346	&	0.364	&	0.011	&	0.551	\\
17.5	&	0.1974	&	1.5079	&	0.1216	&	0.0083	&	0.0790	&	1.78	&	0.247	&	0.368	&	0.016	&	0.445	\\
18.1	&	0.1959	&	1.6296	&	0.1493	&	0.0136	&	0.1044	&	1.67	&	0.326	&	0.469	&	0.026	&	0.522	\\
19.0	&	0.2133	&	1.9239	&	0.1239	&	0.0108	&	0.1426	&	1.59	&	0.277	&	0.508	&	0.020	&	0.490	\\
20.1	&	0.2426	&	2.1250	&	0.0956	&	0.0101	&	0.3170	&	0.37	&	0.495	&	0.835	&	0.019	&	0.737	\\
20.7	&	0.2278	&	2.4074	&	0.1343	&	0.0132	&	0.2163	&	1.53	&	0.396	&	0.683	&	0.025	&	0.624	\\
21.4	&	0.2670	&	2.4646	&	0.1322	&	0.0120	&	0.0954	&	1.75	&	0.299	&	0.420	&	0.023	&	0.566	\\
25.4	&	0.3765	&	3.5207	&	0.1051	&	0.0119	&	0.4236	&	0.28	&	0.540	&	1.078	&	0.022	&	0.917	\\
25.9	&	0.3518	&	3.7937	&	0.1491	&	0.0179	&	0.4278	&	0.40	&	0.771	&	1.159	&	0.034	&	1.122	\\
26.3	&	0.4097	&	3.7250	&	0.1423	&	0.0157	&	0.2861	&	0.57	&	0.730	&	0.845	&	0.030	&	1.140	\\
27.2	&	0.4449	&	4.0087	&	0.1661	&	0.0188	&	0.2071	&	0.92	&	0.650	&	0.714	&	0.035	&	1.095	\\
60.0	&	0.7750	&	3.4600	&	0.1234	&	0.0131	&	0.1071	&	1.31	&	0.336	&	0.430	&	0.025	&	1.111	
\enddata
\tablecomments{M$\rm_{*}$ represent the mass of the progenitor star. The values $\rm M_{i}$ represents the total mass of the given atom $i$ \citep{woosley07,sukhbold16}. $\rm N_{Mg}/N_{Si}$ is the ratio of the numbers of Mg to Si atoms in the ejecta. The dust masses $\rm M_{d}$ represent the maximum masses of dust that can form in the ejecta of a given composition, assuming a 100 \% condensation efficiency. The amount of $\rm Mg_{0.7}SiO_{2.7}$ is limited by the total number of Mg or Si atoms, depending on whether the Mg/Si number ratio is less or greater than 0.7. $\rm M_{d}$ (MgO+SiO$_2$) is the maximum amount of dust that can form, assuming that all Mg and Si are locked up in MgO and SiO$_2$ grains. $\rm M_{d}$ ($\rm Al_2O_3$) assumes all Al is locked up in dust, and $\rm M_{d}$ ($\rm Mg_{0.7}SiO_{2.7}+C$) is the combined mass of the maximum possible $\rm Mg_{0.7}SiO_{2.7}$ and carbon dust yields.}
\end{deluxetable*}


\subsection{Mass} \label{mass}

The dust mass surface density map in the right panel of Figure~\ref{dustmaps} shows that the dust in the G54.1+0.3 is fairly uniformly distributed throughout the shell, with one region of enhancement corresponding to Region~1 of Figure~\ref{3color}. The dust mass associated with this enhancement is approximately $\sim$~5\% of the total dust mass in the shell. We note that this region does not coincide with the temperature peak at the western edge of the PWN where the gas density enhancement is found by \citet{temim10}, but instead seems to spatially coincide with the [\ion{Si}{2}] peak seen in Figure~11 of \citet{temim10}. The emission from this region was suggested to arise from ejecta material swept-up by the PWN \citep[see Figure~14 of][]{temim10}.

The best-fit total dust masses for various composition combinations are listed in Table~\ref{tab2}. We present the results that assume the long-wavelength emission originates from the shell, as well as the results that treat this emission as background emission. The models for which the mid-IR emission arises solely from $\rm Mg_{0.7}SiO_{2.7}$ grains produce unreasonably large best-fit dust masses that are inconsistent with nucleosynthetic yields for SN ejecta. This suggests that other grain species with higher emissivities at longer wavelengths are also present in the shell. When a secondary dust composition is included in the models, the best-fit dust masses are somewhat lower, but still higher than what might be expected for SN-condensed dust. For example, for models considering carbon, silicate, or alumina grains as secondary components, the best-fit dust masses range from 1.1-1.5~$\rm M_{\odot}$. However, since the temperature and mass of the secondary component are poorly constrained, the uncertainty on the total mass can be as high as $\pm1$~$\rm M_{\odot}$. The lower dust mass limit for the best fit models in Table~\ref{tab2} (the best-fit value minus the uncertainty) ranges from 0.26 to 1.0 $\rm M_{\odot}$, depending on the chosen composition of the secondary component. These lower limits are achieved when the flux density contribution of the secondary dust component to mid-IR wavelengths is at a maximum, which slightly raises the temperature and decreases the normalization of the primary $\rm Mg_{0.7}SiO_{2.7}$ dust component, effectively decreasing the total dust mass in the shell.

We emphasize that the inferred dust masses are highly dependent on the model that we use to fit the SED. While our choice of grain compositions was motivated by the observed spectral feature and most common dust species arising from dust condensation models, the relative contributions and temperatures of the various dust component will need to be verified by physical dust heating models that properly take into account the radiation field produced by the stellar cluster and a distribution of dust grain sizes and temperatures. For example, a model that included radiative heating of dust grains with a continuous size distribution in the Crab Nebula resulted in a total dust mass that is a factor of two lower than for a two-temperature model that fits the data equally well and uses the same dust grain composition \citep{temim13}. In future work, it will be important to determine whether the parameter space produced by a heating model will allow for a significantly lower total dust mass in G54.1+0.3. Additional limitations and sources of uncertainty are discussed in Section~\ref{caveats}.

\subsection{Caveats and Limitations}\label{caveats}

The estimated dust masses in this work are highly dependent on the chosen model parameters and would vary with changes in the source distance and dust grain composition, shape, size distribution, porosity, and clumping. One of the main assumptions that drives the total dust mass is that the 21~\micron\ feature and a significant fraction of the mid-IR continuum in the spectrum of the G54.1+0.3 shell arise from $\rm Mg_{0.7}SiO_{2.7}$ grains. Based on the available optical constants, this grain composition is the only one that reproduces the precise shape of the 21~\micron\ feature \citep[see][]{arendt14}. However, there may be other grain compositions with higher emissivities at wavelengths longward of this feature that reproduce it equally well, which would in turn lower the estimated dust mass.

As mentioned in Section~\ref{mass}, introducing a wide distribution of grain sizes, and therefore temperatures, may also affect the relative contribution of the various dust species to the global SED and change the derived dust mass. The dust mass would also change if the grains are porous or not spherical. While we cannot rule out porous grains in our model, studies of meteoritic grains that likely originate from SNe show that they are compact \citep{molster10}. Since compact grains were also used to estimate the the dust mass in of SN 1987A, for example, our estimate for G54.1+0.3 at least provides a relative dust mass in relation to those estimated for other SNRs that contain SN-condensed dust.

Another matter of concern is the balance of heating and cooling of the dust in the shell. The IR observations show that only a small fraction of the stellar radiation is reradiated by the dust in the shell. While the available data on the UV absorption properties of $\rm Mg_{0.7}SiO_{2.7}$ grains in particular are incomplete, silicate grains in general are opaque at these wavelengths, so we would expect a much higher fraction of the stellar flux to be absorbed if the total dust mass exceeded a few tenths of solar masses.
However, if the $\rm Mg_{0.7}SiO_{2.7}$ grains are indeed as large as 1~\micron\ (see Section~\ref{temp}), their effective absorption per unit mass will be smaller, so the larger dust masses would likely not be in conflict with the inferred low value of the UV optical depth.
A more detailed physical heating model will be necessary to resolve these ambiguities and determine whether including these additional effects will significantly alter the dust mass estimate for G54.1+0.3.

\section{Discussion}\label{disc}

The fact that the dust in the G54.1+0.3 IR shell shows a fairly uniform spatial mass distribution and no evidence for enhancements at the locations of the stellar sources confirms that the point-like 24~\micron\ excess (Figure~\ref{3color}) does not originate from dust intrinsic to the stars. The dust mass surface density map is consistent with the scenario in which the stars are heating a shell of dust that condensed from the SN ejecta.
The lower mass limit of at least 0.26 $\rm M_{\odot}$ of SN-formed dust derived from the SED fits would be the second largest observationally confirmed dust mass after SN~1987A \citep{matsuura15, dwek15}, assuming that the grains are compact and that the $\rm Mg_{0.7}SiO_{2.7}$ grains are responsible for the observed 21 \micron\ feature and a significant fraction of the mid-IR continuum (additional caveats are discussed in Section~\ref{caveats}). A large dust mass of $\sim$ 0.8 $\rm M_{\odot}$ requiring a high condensation efficiency was also inferred for Cas~A, accounting for the fact that some of the dust has already been destroyed by the reverse shock \cite{micelotta16}. The comparison of the estimated dust mass in G54.1+0.3 with nucleosynthetic yields implies a similarly high condensation efficiency.

The PWN in G54.1+0.3 has not yet been overtaken by the SNR reverse shock, so it is not unexpected that the mass of the SN-formed dust is still high. Just how much of the observed dust will eventually survive the shock and be injected into the ISM remains uncertain. \citet{micelotta16} estimate that only 12--16\% of dust grains will survive the passage of the reverse shock in Cas A. However, this percentage may be higher in the case of Type~IIP progenitors that are expected to form larger grains that are not as easily destroyed. For example, in their models that account for the for the gas-phase chemistry, nucleation, and coagulation of grains, \citet{sarangi15} find that Type~IIP SNe tend to form larger grains, especially if the ejecta are clumpy, and that ejecta clumpiness also leads to somewhat higher dust masses and a higher fraction of metallic grains. In addition, if the SN that produced G54.1+0.3 occurred in a low density bubble produced by the OB association, the reverse shock may be very weak once it encounters the SN-condensed dust that surrounds the PWN, leading to a higher survival rate of grains and a higher eventual dust mass input into the ISM. This hypothesis will need to be tested with more detailed modeling of the dust destruction by the SN reverse shock.

In order to compare our estimated dust masses with nucleosynthetic yields, we list the yields of C, O, Mg, Al, and Si for different progenitor masses in Table~\ref{tab3}, based on the models of \citet{woosley07} and \citet{sukhbold16}. The table also lists the maximum possible mass of $\rm Mg_{0.7}SiO_{2.7}$, MgO + SiO$_2$,  $\rm Al_2O_3$, and carbon dust that could form in the ejecta assuming a 100\% condensation efficiency. The amount of $\rm Mg_{0.7}SiO_{2.7}$ is limited by the total number of the Mg or Si atoms, depending on whether $\rm N_{Mg}/N_{Si}$ is less or greater than 0.7.
The mass of MgO+SiO2 assumes that all Mg and Si in the ejecta are locked up in MgO and SiO$_2$ and/or one or more forms of $\rm Mg_xSiO_{2+x}$ grains. We compare these values to the lower dust mass limit in the G54.1+0.3 shell, listed in Table~\ref{tab2}. The lower mass limit of $\sim$ 0.25 $\rm M_{\odot}$ of $\rm Mg_{0.7}SiO_{2.7}$ grains is consistent with nucleosynthetic constraints for a progenitor of at least $\sim$~15~$\rm M_{\odot}$, but this requires a $\rm Al_2O_3$  mass of $\sim$~0.03 $\rm M_{\odot}$, which corresponds to a progenitor mass of at least 21 $\rm M_{\odot}$ (see Table~\ref{tab3}). However, slightly increasing the mass of $\rm Mg_{0.7}SiO_{2.7}$ would decrease the required mass of $\rm Al_2O_3$ and therefore lower this limit on the progenitor mass. The next lowest dust mass lower limit of 0.28 $\rm M_{\odot}$ is for the $\rm Mg_{0.7}SiO_{2.7}$ plus carbon dust model, and implies a progenitor mass of at least 16 $\rm M_{\odot}$. 

The primary composition of the grains may also offer some clues about the SN progenitor. For example, the $\rm Mg_{0.7}SiO_{2.7}$ grains that make up a large fraction of the dust are characterized by a Mg to Si ratio of 0.7. The sixth column of Table~\ref{tab3} lists the ratio of the numbers of Mg to Si atoms in the ejecta ($\rm N_{Mg}/N_{Si}$) for different progenitor masses. This ratio is less than unity for progenitor masses of $\sim$~27~$\rm M_{\odot}$ and below. A ratio below unity may explain the formation of the less common magnesium silicate grain species, such as $\rm Mg_{0.7}SiO_{2.7}$, instead of the more common ones like forsterite (Mg$_2$SiO$_4$) for which Mg/Si = 2. The mass and the composition of the grains in G54.1+0.3 therefore suggest that the mass of the SN progenitor was in the 16--27~$\rm M_{\odot}$ range, consistent with the estimate of \citet{gelfand15} and the range of 18--35~$\rm M_{\odot}$ suggested by the analysis of the stellar cluster by \citet{kim13}. We note here that a dust condensation efficiency of $<$100\% would lead to a more massive progenitor.

\section{Conclusions}\label{conclusions}

We analyzed the 15--500 \micron\ IR emission from a dusty shell of SN-formed dust surrounding the PWN in the SNR G54.1+0.3. We find that the SED, and in particular a spectral feature at 21 $\mu$m, is well described by a dust composition of $\rm Mg_{0.7}SiO_{2.7}$ grains, with a secondary component possibly arising from carbon, silicate, or alumina grains. Through a spatially resolved analysis of the IR emission, we derive dust temperature and mass surface density maps that confirm the scenario in which stellar members of the SN progenitor's cluster are the primary heating sources for the SN dust. The smallest total dust mass resulting from our models is 1.1~$\pm$~0.8~$\rm M_{\odot}$, assuming compact grains and a distance of 6~kpc. Self-consistent radiative heating models that invoke a continues distribution of grain sizes may affect this estimate and allow for a lower mass.
Nevertheless, the large quantity of dust inferred from our model implies a high dust condensation efficiency, as has been suggested for both SN~1987A and Cas~A \citep{matsuura15, dwek15,micelotta16}. A comparison of the dust mass and composition with nucleosynthetic yields suggests that G54.1+0.3 resulted from a 16--27~$\rm M_{\odot}$ progenitor. Since the dusty shell has not yet been encountered by the SN reverse shock, the ultimate survival of the dust remains unclear. This study implies that dust can efficiently form in the ejecta of Type~IIP SN explosions, and that certain classes of SNe may indeed be significant sources of dust in the Universe. Future high spatial resolution imaging and spectroscopic observations, particularly with the \textit{James Webb Space Telescope}, as well as detailed physical dust heating models will be necessary to confirm these results.

\

\acknowledgments

This work is based in part on observations made with \textit{Herschel}. \textit{Herschel} is an ESA space observatory with science instruments provided by European-led Principal Investigator consortia and with important participation from NASA.
This work is based in part on observations made with the \textit{Spitzer Space Telescope}, which is operated by the Jet Propulsion Laboratory, California Institute of Technology, under a contract with NASA.
This research includes observations with AKARI, a JAXA project with the participation of ESA.
Based (in part) on observations made with the NASA/DLR Stratospheric Observatory for Infrared Astronomy (SOFIA). SOFIA is jointly operated by the Universities Space Research Association, Inc. (USRA), under NASA contract NAS2-97001, and the Deutsches SOFIA Institut (DSI) under DLR contract 50 OK 0901 to the University of Stuttgart. We acknowledge financial support for this work that was provided by NASA through award SOF \#04-0167 issued by USRA.

\bibliographystyle{apj}

\end{document}